\documentclass[10pt,aps,prd,amsmath,amssymb,twoside,showpacs, nofootinbib,preprintnumbers,notitlepage]{revtex4-1}

\usepackage[T1]{fontenc}
\usepackage{fix-cm}
\usepackage{ifthen}
\usepackage{dsfont} % replaces bbm to avoid type 3 fonts
\usepackage{upgreek}
\usepackage[dvipsnames,svgnames]{xcolor}
\usepackage[citebordercolor=Apricot,% 	citations
				linkbordercolor=Apricot,% 		equations,figures,sections,etc.
				urlbordercolor=Apricot% 	URLs, emails,etc.
				]{hyperref}

\newcommand{\nc}{\newcommand}

\nc{\defeq}{:=}
\nc{\im}{\mathrm{i}}  	% imaginary i
\nc{\xd}{\mathrm{d}}	 	% differential d 

\nc{\mc}[1]{\mathcal{#1}}
\nc{\xD}{\mc{D}} 	% path integral measure
\nc{\cH}{\mc{H}} 	% Hilbert space

\nc{\sphere}[1][]{\mathds{S}^{#1}}
\nc{\reals}[1][]{{\mathds{R}^{#1}}}
\nc{\complex}[1][]{{\mathds{C}^{#1}}}

\nc{\Repart}{\mathds{R}\text{e}\:}
\nc{\Impart}{\mathds{I}\text{m}\:}

\nc{\change}[1]{\textcolor{Red}{\bfseries{#1}}}

\nc{\unemptynumberone}[2]{\ifthenelse{\equal{#1}{}}{}{#2}}

\nc{\mcW}[5]{{\mc W}^{#1#2}(#3_{#4},#3_{#5})}
\nc{\AR}{A^{\reals}}
\nc{\KS}{\mathcal{K}^\mathrm{S}}
\nc{\KD}{\mathcal{K}^\mathrm{D}}

\nc{\bal}[1]{\begin{align}#1\end{align}}
\nc{\bals}[1]{\begin{align*}#1\end{align*}}
\nc{\balsplit}[1]{\bal{\begin{split}#1\end{split}}}

\nc{\vc}[1]{\underline{#1}}
\nc{\lvc}[2][]{_{#1\vc #2}}
\nc{\ltx}[1]{_{\text{#1}}}
\nc{\htx}[1]{^{\text{#1}}}

\nc{\intval}[3]{[#1_{#2},#1_{#3}]}
\nc{\lintval}[3]{ _{ [#1_{#2},#1_{#3}] }}

\nc{\ti}{\tilde}

\nc{\del}{\partial}
\nc{\deth}[1][]{\det\!^{#1}\,\!}

\nc{\eu}{\mathrm{e}}		% Euler number e = 2.71
\nc{\coco}{\overline}		% complex conjugation

\nc{\Id}[1][]{{\text{Id} \ltx{#1} \,\!}}
\nc{\One}[1][]{{\mathds{1} \ltx{#1} \,\!}}

\nc{\sign}{\,\text{sign}\,}

\nc{\cHS}{\cH^{\mathrm{S}}}

\nc{\psistate}[3][]{\ps^{{#3}\unemptynumberone{#1}{,}#1}_{#2}\,\!}
\nc{\psS}[2][]{\psistate[#1]{#2}{\text{S}}}	% Schr\"odinger
\nc{\psSar}[3][]{\psS[#1]{#2}(#3)}
\nc{\psD}[2][]{\psistate[#1]{#2}{\text{D}}}	% Dirac
\nc{\psDar}[3][]{\psD[#1]{#2}(#3)}
\nc{\psH}[2][]{\psistate[#1]{#2}{\text{H}}}	% Holomorphic
\nc{\psHar}[3][]{\psH[#1]{#2}(#3)}

\nc{\roS}[2][]{\ro^{\text{S}\unemptynumberone{#1}{,}#1}_{#2}}
\nc{\roH}[2][]{\ro^{\text{H}\unemptynumberone{#1}{,}#1}_{#2}}

\nc{\Sibar}{{\overline\Si}}

\nc{\Norm}[2][]{\mc N^{#1}_{#2}}

\nc{\bgl}{\left}
\nc{\biigl}{\Bigl}
\nc{\biiigl}{\biggl}
\nc{\biiiigl}{\Biggl}

\nc{\bgr}{\right}
\nc{\biigr}{\Bigr}
\nc{\biiigr}{\biggr}
\nc{\biiiigr}{\Biggr}

\nc{\bgm}{\middle}
\nc{\biigm}{\Bigm}
\nc{\biiigm}{\biggm}
\nc{\biiiigm}{\Biggm}

\nc{\bglrr}[1]{\bgl( #1 \bgr)}
\nc{\biglrr}[1]{\bigl( #1 \bigr)}
\nc{\biiglrr}[1]{\biigl( #1 \biigr)}
\nc{\biiiglrr}[1]{\biiigl( #1 \biiigr)}
\nc{\biiiiglrr}[1]{\biiiigl( #1 \biiiigr)}

\nc{\bglrs}[1]{\bgl[ #1 \bgr]}
\nc{\biglrs}[1]{\bigl[ #1 \bigr]}
\nc{\biiglrs}[1]{\biigl[ #1 \biigr]}
\nc{\biiiglrs}[1]{\biiigl[ #1 \biiigr]}
\nc{\biiiiglrs}[1]{\biiiigl[ #1 \biiiigr]}

\nc{\bglrc}[1]{\bgl\{ #1 \bgr\}}
\nc{\biglrc}[1]{\bigl\{ #1 \bigr\}}
\nc{\biiglrc}[1]{\biigl\{ #1 \biigr\}}
\nc{\biiiglrc}[1]{\biiigl\{ #1 \biiigr\}}
\nc{\biiiiglrc}[1]{\biiiigl\{ #1 \biiiigr\}}

\nc{\abs}[1]{\left| #1 \right|}
\nc{\abss}[1]{\bigl| #1 \bigr|}
\nc{\absss}[1]{\biigl| #1 \biigr|}
\nc{\abssss}[1]{\biiigl| #1 \biiigr|}
\nc{\absssss}[1]{\biiiigl| #1 \biiiigr|}

\nc{\fracwspace}{\hspace{0.3ex}}
\nc{\fracw}[2]{\, \frac{\fracwspace #1 \fracwspace}%
			{\fracwspace #2 \fracwspace} \,%
			}
\nc{\tfracw}[2]{\, \tfrac{\fracwspace #1 \fracwspace}%
			{\fracwspace \phantom{X^X}\hspace{-3ex} #2 \fracwspace} \,%
			}

\nc{\bosym}{\boldsymbol}

\newlength{\bralength}
\newlength{\brasymlength}
\newlength{\braasymlength}
\newlength{\braaasymlength}
\newlength{\braaaasymlength}
\newlength{\braaaaasymlength}
\nc{\bra}[2][]{{ \settowidth{\bralength}{$_{#1}$} \hspace{\bralength}
					\bosym{\langle}_{\,\hspace{-\bralength}\hspace{-\brasymlength}{#1}\hspace{\brasymlength}}
					#2 \,\bosym{|}
				}}
\nc{\braa}[2][]{ \settowidth{\bralength}{$_{#1}$} \hspace{\bralength}
				\bosym{\bigl\langle}_{\hspace{-\bralength}\hspace{-\braasymlength}{#1}\hspace{\braasymlength}}
				 #2 \,\bosym{\bigr|} }
\nc{\braaa}[2][]{ \settowidth{\bralength}{$_{#1}$} \hspace{\bralength}
				\bosym{\biigl\langle}_{\hspace{-\bralength}\hspace{-\braaasymlength}{#1}\hspace{\braaasymlength}}
				 #2 \,\bosym{\biigr|} }
\nc{\braaaa}[2][]{ \settowidth{\bralength}{$_{#1}$} \hspace{\bralength}
				\bosym{\biiigl\langle}_{\,\hspace{-\bralength}\hspace{-\braaaasymlength}{#1}\hspace{\braaaasymlength}}
				 #2 \,\bosym{\biiigr|} }
\nc{\braaaaa}[2][]{ \settowidth{\bralength}{$_{#1}$} \hspace{\bralength}
				\bosym{\biiiigl\langle}_{\:\hspace{-\bralength}\hspace{-\braaaaasymlength}{#1}\hspace{\braaaaasymlength}}
				 #2 \,\bosym{\biiiigr|} }

\nc{\braout}[1]{\bra[\text{out}]{#1}}
\nc{\braaout}[1]{\braa[\text{out}]{#1}}
\nc{\braaaout}[1]{\braaa[\text{out}]{#1}}
\nc{\braaaaout}[1]{\braaaa[\text{out}]{#1}}
\nc{\braaaaaout}[1]{\braaaaa[\text{out}]{#1}}

\nc{\ket}[2][]{{\bosym{|}\, #2 \bosym{\rangle}_{#1} }}
\nc{\kett}[2][]{\bosym{\bigl|}\, #2 \bosym{\bigr\rangle}_{#1} }
\nc{\kettt}[2][]{\bosym{\biigl|}\, #2 \bosym{\biigr\rangle}_{\!#1} }
\nc{\ketttt}[2][]{\bosym{\biiigl|}\, #2 \bosym{\biiigr\rangle}_{\!\!#1} }
\nc{\kettttt}[2][]{\bosym{\biiiigl|}\, #2 \bosym{\biiiigr\rangle}_{\!\!#1} }

\nc{\ketin}[1]{\ket[\text{in}]{#1}}
\nc{\kettin}[1]{\kett[\text{in}]{#1}}
\nc{\ketttin}[1]{\kettt[\text{in}]{#1}}
\nc{\kettttin}[1]{\ketttt[\text{in}]{#1}}
\nc{\ketttttin}[1]{\kettttt[\text{in}]{#1}}

\nc{\ketout}[1]{\ket[\text{out}]{#1}}
\nc{\kettout}[1]{\kett[\text{out}]{#1}}
\nc{\ketttout}[1]{\kettt[\text{out}]{#1}}
\nc{\kettttout}[1]{\ketttt[\text{out}]{#1}}
\nc{\ketttttout}[1]{\kettttt[\text{out}]{#1}}

\nc{\braket}[3][]{{\bosym{\langle} #2 \,\bosym{|}\, #3 \bosym{\rangle}_{#1} }}
\nc{\brakett}[3][]{\bosym{\bigl\langle} #2 \,\bosym{\bigl|}\, #3 \bosym{\bigr\rangle}_{#1} }
\nc{\brakettt}[3][]{\bosym{\biigl\langle} #2 \,\bosym{\biigl|}\, #3 \bosym{\biigr\rangle}_{\!#1} }
\nc{\braketttt}[3][]{\bosym{\biiigl\langle} #2 \,\bosym{\biiigl|}\, #3 \bosym{\biiigr\rangle}_{\!\!#1} }
\nc{\brakettttt}[3][]{\bosym{\biiiigl\langle} #2 \,\bosym{\biiiigl|}\, #3 \bosym{\biiiigr\rangle}_{\!\!#1} }

\nc{\braopket}[4][]{{\bosym{\langle} #2 \,\bosym{|}\, #3 \,\bosym{|}\, #4 \bosym{\rangle}_{#1} }}
\nc{\braopkett}[4][]{\bosym{\bigl\langle} #2 \,\bosym{\bigr|}\, #3 \,\bosym{\bigl|}\, #4 \bosym{\bigr\rangle}_{#1}}
\nc{\braopkettt}[4][]{\bosym{\biigl\langle} #2 \,\bosym{\biigr|}\, #3 \,\bosym{\biigl|}\, #4 \bosym{\biigr\rangle}_{\!#1}}
\nc{\braopketttt}[4][]{\bosym{\biiigl\langle} #2 \,\bosym{\biiigr|}\, #3 \,\bosym{\biiigl|}\, #4 \bosym{\biiigr\rangle}_{\!\!#1}}
\nc{\braopkettttt}[4][]{\bosym{\biiiigl\langle} #2 \,\bosym{\biiiigr|}\, #3 \,\bosym{\biiiigl|}\, #4 \bosym{\biiiigr\rangle}_{\!\!#1}}

\nc{\expval}[2][]{{\bosym{\langle} #2 \bosym{\rangle}_{#1}} }
\nc{\expvall}[2][]{\bosym{\bigl\langle} #2 \bosym{\bigr\rangle}_{#1}}
\nc{\expvalll}[2][]{\bosym{\biigl\langle} #2 \bosym{\biigr\rangle}_{\!#1}}
\nc{\expvallll}[2][]{\bosym{\biiigl\langle} #2 \bosym{\biiigr\rangle}_{\!#1}}
\nc{\expvalllll}[2][]{\bosym{\biiiigl\langle} #2 \bosym{\biiiigr\rangle}_{\!\!#1}}

\nc{\inpro}[3][]{\bosym{\langle} #2 \bosym{,} \, #3 \bosym{\rangle}_{#1}}
\nc{\inproo}[3][]{\bosym{\bigl\langle} #2 \bosym{,} \; #3 \bosym{\bigr\rangle}_{#1}}
\nc{\inprooo}[3][]{\bosym{\biigl\langle} #2 \bosym{,} \;\, #3 \bosym{\biigr\rangle}_{\!#1}}
\nc{\inproooo}[3][]{\bosym{\biiigl\langle} #2 \bosym{,} \;\, #3 \bosym{\biiigr\rangle}_{\!#1}}
\nc{\inprooooo}[3][]{\bosym{\biiiigl\langle} #2 \bosym{,} \;\, #3 \bosym{\biiiigr\rangle}_{\!\!#1}}

\nc{\al}{{\alpha}}
\nc{\be}{{\beta}}
\nc{\ga}{{\gamma}}
\nc{\de}{{\delta}}
\nc{\ep}{{\epsilon}}
\nc{\vep}{{\varepsilon}}
\nc{\ph}{{\phi}}
\nc{\vph}{{\varphi}}
\nc{\ps}{{\psi}}
\nc{\et}{{\eta}}
\nc{\io}{{\iota}}
\nc{\ka}{{\kappa}}
\nc{\la}{{\lambda}}
\nc{\ro}{{\rho}}
\nc{\si}{{\sigma}}
\nc{\ta}{{\tau}}
\nc{\te}{{\theta}}
\nc{\vte}{{\vartheta}}
\nc{\om}{{\omega}}
\nc{\ki}{{\chi}}
\nc{\ze}{{\zeta}}

\nc{\Ga}{{\Gamma}}
\nc{\De}{{\Delta}}
\nc{\Ph}{{\Phi}}
\nc{\Ps}{{\Psi}}
\nc{\La}{{\Lambda}}
\nc{\Si}{{\Sigma}}
\nc{\Om}{{\Omega}}
\nc{\Te}{{\Theta}}
\nc{\Up}{{\Upsilon}}

\nc{\upal}{{\upalpha}}
\nc{\upbe}{{\upbeta}}
\nc{\upga}{{\upgamma}}
\nc{\upde}{{\updelta}}
\nc{\upep}{{\upepsilon}}
\nc{\upvep}{{\upvarepsilon}}
\nc{\upph}{{\upphi}}
\nc{\upvph}{{\upvarphi}}
\nc{\upps}{{\uppsi}}
\nc{\upet}{{\upeta}}
\nc{\upio}{{\upiota}}
\nc{\upka}{{\upkappa}}
\nc{\upla}{{\uplambda}}
\nc{\upro}{{\uprho}}
\nc{\upsi}{{\upsigma}}
\nc{\upta}{{\uptau}}
\nc{\upte}{{\uptheta}}
\nc{\upvte}{{\upvartheta}}
\nc{\upom}{{\upomega}}
\nc{\upki}{{\upchi}}
\nc{\upze}{{\upzeta}}

\usepackage{geometry}
\begin{document}

% ================================================================
\title{The S-matrix in Schr{\"o}dinger Representation
		for Curved Spacetimes\\ 
		in General Boundary Quantum Field Theory}
\author{Daniele Colosi}
\email{dcolosi@enesmorelia.unam.mx}
\author{Max Dohse}
\email{max@ifm.umich.mx}
\affiliation{$^*$Escuela Nacional de Estudios Superiores, 
			Unidad Morelia,\\ 
			Universidad Nacional Aut\'onoma de M\'exico (UNAM),\\ 
			Campus Morelia, C.P.~58190, Morelia, Mexico
				\\
			$^\dagger$Instituto de F\'isica y Matem\'aticas 
				(IFM-UMSNH),\\
				Universidad Michoacana de San Nicol\'as de Hidalgo,\\
				Edificio C-3, Ciudad Universitaria, 
				58040 Morelia, M\'exico.
			}

% ================================================================

\begin{abstract}
	\noindent%
	We use the General Boundary Formulation (GBF) of Quantum Field Theory
	to compute the S-matrix for a general interacting scalar field
	in a wide class of curved spacetimes.
	As a by-product we obtain the general expression 
	of the Feynman propagator for the scalar field, 
	defined in the following three types of spacetime regions.
	First, there are the familiar interval regions
	(e.g.~a time interval times all of space).
	Second, we consider the rod hypercylinder regions
	(all of time times a solid ball in space).
	Third, the tube hypercylinders 
	(all of time times a solid shell in space)
	are related to interval regions, and result from 
	removing a smaller rod from a concentric larger one.
	Using the Schr{\"o}dinger representation for the quantum states
	combined with Feynman's path integral quantization,
	we obtain the S-matrix as the asymptotic limit of the GBF amplitude
	associated with finite interval and rod regions.
	For interval regions, whose boundary consists of two Cauchy surfaces,
	the asymptotic GBF-amplitude becomes the standard S-matrix.
	Our work generalizes previous results (obtained in Minkowski, Rindler,
	de Sitter, and Anti de Sitter spacetimes)
	to a wide class of curved spacetimes.
\end{abstract}

% ================================================================

\maketitle

\hspace{20mm} General Boundary Formulation; Feynman propagator

%
% ================================================================
%
% practical for Arxiv version and referees, 
% can be removed later by journal
\tableofcontents
%
% ================================================================
%
\section{Introduction}
\label{sec:intro}
\noindent%
The purpose of this paper is the derivation of the general structure of the scattering matrix for a quantum scalar field defined on a broad class of flat and curved spacetimes. In the usual treatment, the S-matrix refers to scattering processes for states defined on (asymptotic) spacelike Cauchy surfaces. The results we present generalize this situation to the case of states defined on certain classes of timelike hypersurfaces. 
S-matrices of this new type can be computed explicitly, the corresponding Feynman rules of perturbation theory can be derived, and an appropriate notion of probability can be extracted from them, by adopting the General Boundary Formulation (GBF) \cite{Oe:boundary,Oe:GBQFT,Oe:KGtl,Oe:probgbf} of Quantum Field Theory (QFT). The novelty of the GBF resides firstly in associating Hilbert spaces of quantum states to {\em arbitrary} hypersurfaces (of codimension one) in spacetime. Cauchy surfaces are then only a special choice, not obligatory within the GBF. In particular, the boundaries of spacetime regions (regions have codimension zero) are hypersurfaces, and hence have their associated quantum state spaces. Secondly, amplitudes are associated with spacetime regions and are determined by a linear map from the state spaces on their boundaries to the complex numbers. These algebraic structures (state spaces and amplitudes) are required to satisfy a set of axioms that guarantees their coherence. Finally, the GBF provides a consistent probabilistic interpretation for the regions' amplitudes, which generalizes Born's rule.

The new types of scattering matrices shed light on geometrical aspects of QFT.
Moreover, the extension provided by the GBF is a necessity in situations where the usual S-matrix fails for some reason, as for example in Anti-de Sitter spacetime where no asymptotic temporal regions exist. 
Another situation where the GBF is expected to offer the appropriate tools is in describing the dynamics of fields in the presence of an eternal black hole: 
No free temporal asymptotic states can be defined since the interaction of a massive quantum field with a black hole will never vanish in time. 
However, far away from the black hole, a notion of free {\em spatial} asymptotic states is available, and one can compute amplitudes for these states within the GBF. 

Our main goal here is to contribute to the development of the GBF. 
Indeed this work can be seen as a generalization of previous results obtained in Minkowski \cite{CoOe:letter,CoOe:smatrix}, Euclidean \cite{CoOe:2d}, Rindler \cite{CoRa:rindler} and de Sitter spacetimes \cite{Co:letter,Co:dS}. Inspired by these papers, we will consider two classes of spacetime regions. The first class is characterized by a boundary consisting of several disjoint hypersurfaces, which we do not require to be Cauchy or even spacelike. Apart from the metric nature of the hypersurfaces involved, this is close to the usual time-interval region of spacetime, whose boundary consists of one initial and one final equal-time hypersurface (respectively Cauchy surfaces). The dynamics taking place in these regions can hence be understood as the evolution of a quantum state from an initial hypersurface to a final one. The second class is radically different: Here, the boundary is completely connected and timelike, two aspects not treatable within the standard formulation of QFT. The dynamics take place inside the region enclosed by the boundary.

The outline of the paper is as follows. In the next section we introduce the two types of spacetime regions we will be interested in. We describe there the classical theory of a real scalar field, expressing the solution of the Klein-Gordon equation in terms of boundary field configurations. In Section \ref{sec:quantum}, the Schr\"odinger-Feynman quantization prescription is described in the GBF context, and the main structures corresponding to the different regions considered are defined. In Section \ref{sec:freetheory} the quantum amplitudes for states of the free theory are computed. This result is obtained in three steps: First, we evaluate the path integral of the field propagator for the regions of interest in Section \ref{sec:freeprops}. 
Then, we introduce vacuum and coherent states in Section \ref{sec:vacuum} and Section \ref{sec:coherent} respectively. 
Finally, in Section \ref{sec:freeampl} we obtain the expressions of the free amplitudes. We treat the interacting theory in Section \ref{sec:inttheory}, starting with the interaction of the scalar field with a source field in Sections \ref{sec:intamplinterval} and \ref{sec:intamplhyp}. Subsequently we use functional derivative techniques to obtain the amplitudes for the general interacting theory in Section \ref{sec:genint}. We summarize and discuss our results in Section \ref{sec:discussion}.
%
% ================================================================
%
\section{Classical theory}
\label{sec:_classical}
\noindent%
We consider a real, massive, minimally coupled Klein-Gordon field 
$\phi$ in a 4-dimensional curved spacetime manifold 
with Lorentzian signature and metric tensor $g_{\mu\nu}$. 
We consider orientable spacetimes that admit
at least one global time coordinate (time function).
We also require that spacetime can be foliated
as indicated at the beginning of Section~\ref{sec:interval},
respectively Section~\ref{sec:hypercylinder}.
That is, we only consider spacetimes for which such a foliation exists,
and use only such foliations.
An important class of spacetimes which fulfill these requirements
consists of all globally hyperbolic spacetimes: 
They can be foliated by Cauchy surfaces
with the foliation parameter being a time function,
and the metric splits with respect to this time function
(becomes block diagonal) \cite{BernalSanchez:_MetricSplitting}.
However, we shall not require spacetime to be globally hyperbolic,
because this would exclude several interesting spacetimes,
for example Anti de Sitter and black hole spacetimes
(Schwarzschild is globally hyperbolic, whereas Kerr is not).
Another important requirement is that the foliation must allow 
separation of variables (with respect to the foliation parameter)
in the Klein-Gordon equation, see also Sections~\ref{sec:interval}
and \ref{sec:hypercylinder}.

We introduce the free action $S^0_{M}$
in a spacetime region $M$ via the following bilinear form,
which is symmetric due to the symmetry of the metric tensor $g_{\mu\nu}$. 
Denoting the chosen time coordinate by $y^0$, and defining 
$\si_{00} \defeq \sign g_{00}$ (making the expression independent 
of the metric's overall sign), the free action is defined by
\bal{		% clear
	\label{eq:action}
	\tilde S^0_{M}(\eta,\zeta) 
	& = \frac{1}{2} \int_M \xd^4 y\, \sqrt{|g|}\, 
		\Bigl( \si_{00} g^{\mu \nu} (\del_{y^\mu} \et)\, 
		(\del_{y^\nu} \ze) - m^2\,\et\,\ze \Bigr), 
		\\
	S^0_M(\phi) & =	\tilde S^0_{M}(\phi,\phi),
	}
wherein we integrate over the region $M$ and use the notation 
$\del_{y^\mu} = \del / \del y^{\mu}$.
By $g$ we denote the determinant of the metric tensor: $g \equiv \det g_{\mu \nu}$, and $m$ indicates the mass of the field.
The action's label $0$ refers to the free theory. 
We use Einstein's sum convention in the form that a summation is understood over all greek lowercase indices which appear exactly once as a superscript and once as a subscript in a term.
The variation of the free action yields the (homogeneous) Klein-Gordon equation as the Euler-Lagrange equation of \eqref{eq:action}:
\bal{		% clear
	\label{eq:KG}
	\Bigl(\si_{00} \Box + m^2 \Bigr)\, \phi(y)
	= \Bigl(\si_{00}
	\tfrac{1}{\sqrt{|g|}}\, \del_{y^\mu} \sqrt{|g|}\,
	g^{\mu \nu} \del_{y^\nu} + m^2 \Bigr)\, \phi(y)
	= 0.
	}
If $\et(y)$ and $\ze(y)$ are solutions of this Klein-Gordon equation, 
then after performing an integration by parts in \eqref{eq:action}
we obtain:
\bal{		% clear
	\label{eq:action_solutions}
	\tilde S^0_{M}(\eta,\zeta) 
	& = \frac{1}{2} \int_M \xd^4 y\;\del_{y^\mu} \sqrt{|g|}\, 
		 \si_{00} g^{\mu \nu}\, \et\, \del_{y^\nu} \ze.
	}
Assuming further that $\et,\ze$ are compactly supported on $M$,
we can apply Stokes' Theorem which results in the following boundary integral
(wherein we use adapted coordinates $x=(x^0,\vc x)$ such that $\vc x =(x^1,x^2,x^3)$ are some coordinates on the boundary $\del M$ of $M$ on which $x^0 = \text{const.}$, with now $\del_\mu=\del/\del x^\mu$, $\del_0$ pointing outward, and $\xd^3x :=\xd x^1\,\xd x^2\,\xd x^3$, while $g^{(3)}$ is the determinant of the induced metric on the boundary, and $n_{\mu}$ is the Riemannian outward unit normal on $\del M$, see (13) in \cite{Oe:KGtl}):
\bal{		% revised 2016 MAR 23
	\label{eq:actreg}
	\tilde S^0_{M}(\et,\ze) 
	= \frac{1}{2}
	\si_{00} 
	\int_{\del M} \xd^3 x \sqrt{|g^{(3)}|}\; 
		\et \left( n^{\mu} \del_{\mu} \ze \right)
	= \tilde S^0_{M}(\ze,\et).
	}
The symmetry of this expression is not manifest in the formula,
but is caused by the fact that $\et$ and $\ze$ are Klein-Gordon solutions.
We will be interested in studying the dynamics of the field in two different types of spacetime regions called interval regions and hypercylinder regions respectively. 
The next two subsections are devoted to the definition of these regions and to the expression of the action for the field defined there in terms of the boundary field configurations%
\footnote{
	We shall denote classical Klein-Gordon solutions on spacetime by the 	
	''tall'' letters $\ph,\xi,\ze,\la(\ta,\vc x)$, whereas configurations 
	on hypersurfaces of constant $\ta$ are denoted by the ''short'' letters 
	$\vph, \ki, \et(\vc x)$.
	}% 
\,$\vph$:
\bal{
	% clear
	\label{eq:_ph_vph}
	\vph = \phi\bigr|_{\del M}.
	}
Letting now $\phi$ and $\phi_0$ classical solutions 
that are compactly supported on $M$, 
with $\phi_0$ vanishing on the boundary $\del M$,
using \eqref{eq:actreg} we obtain:
\bal{S^0_M(\phi\!+\!\phi_0)
	& = \tilde S^0_M(\phi\!+\!\phi_0,\phi\!+\!\phi_0)
	= \tilde S^0_M(\phi,\phi)
		+2\tilde S^0_M(\phi_0,\phi) +\tilde S^0_M(\phi_0,\phi_0)
	= \tilde S^0_M(\phi,\phi) = S^0_M(\phi).
	}
That is, the free action $S^0_M(\ph)$ is \emph{uniquely} determined 
by the configuration $\vph = \phi(x)|_{\del M}$ on the boundary. 
This becomes relevant in Section~\ref{sec:freetheory},
where the field propagator $Z^0_M(\vph)$ of a boundary configuration
$\vph$ is calculated via the action $S(\phi)$
of the associated classical solution.
%
% ==============================================================
%
\subsection{Interval regions}
\label{sec:interval}
\noindent%
In this section we will follow the treatment of \cite{CoOe:unit}.
For interval regions $M\lintval\ta12 = [\ta_1,\ta_2] \times I^{(3)}$, 
we suppose a smooth coordinate system $(\ta, \vc x )$ 
which defines a foliation of this region.
We denote by $\ta \in [\ta_1,\ta_2] \subseteq \reals$ the foliation parameter indexing the leaves, while coordinates on the leaves are denoted by $\vc{x}=(x^1,x^2,x^3) \in I^{(3)} \subseteq \reals[3]$. 
It is important to notice that $\ta$ does not necessarily need to be a time variable of the physical theory, and consequently $\vc x $ are not required to be purely spatial coordinates.
The canonical orientation of our leaves is in negative $\ta$-direction, to which we refer as backwards orientation.
We do require the leaves $\Si_\ta$ of constant $\ta$ to be either spacelike in all points or timelike in all points.
We also require the metric to be block diagonal with respect to the foliation, that is $g_{\ta x^i} = 0 = g^{\ta x^i}$ for all $i \in \{1,2,3 \}$.

Each interval region is bounded by two disjoint constant-$\ta$ hypersurfaces: $\Sigma_1$ at $\ta_1$ and $\Sigma_2$ at $\ta_2$. 
Since we orient boundaries as pointing outwards of the enclosed region,
the interval region's boundary can be written as the disjoint union
$\del M\lintval\ta12 = \Si_1\cup\Sibar_2$,
wherein the bar denotes orientation reversal.
In the case that $\ta$ is a time coordinate, and if the boundary consists of Cauchy surfaces, then the interval region is the usual setting for QFT in curved spacetime.
Independently of this, we suppose that our foliation is such that the whole spacetime can be covered by an interval region via sufficiently decreasing $\ta_1$ and increasing $\ta_2$.

The free action of a solution of the Klein-Gordon equation takes the following 
form in interval regions (here and below, the label $[\ta_1,\ta_2]$ indicates 
that the corresponding quantity is computed for an interval region):
\bal{
% clear, revised 2016 MAR 23
S^0_{[\ta_1, \ta_2]} (\phi) 
= \frac{1}{2} \si \! \int\!\! \xd^3 x
	\left\{\! \biiglrr{\!\sqrt{|g^{(3)} g^{\ta \ta}|} \, \phi
					(\del_{\ta} \phi)\!} (\ta_2, \vc x ) \!
			-\! \biiglrr{\!\sqrt{|g^{(3)} g^{\ta \ta}|} \, \phi
				 (\del_{\ta} \phi)\!} (\ta_1, \vc x )  \!
			\right\}\!,
\label{eq:actintreg}
}
wherein $g^{(3)}$ denotes the metric restricted to the hypersurfaces $\Sigma_1$ and $\Sigma_2$ respectively, and $\si \defeq \sign g_{00}\, \sign g^{\ta\ta}$.
For later convenience, it is useful to introduce mode decompositions for the Klein-Gordon solutions and for the boundary field configurations.
With $\vc k $ denoting the set of three parameters (momenta) labeling the modes, we assume that there is a set of complex modes $\{U_{\vc k }(\vc x )\}$ fulfilling the reflection property $U_{\vc k }(\vc x )=\coco{U_{-\vc k }(\vc x )}$, which forms a complete orthonormal basis in the space of field configurations on the hypersurfaces $\Si_\ta$, and also in momentum space:
\bal{% revised 2016 MAR 23
	\label{eq:modenormbasis}
	\int\!\! \xd^3 k\; w_{\vc k}(\vc x)\,
	U_{\vc k }(\vc x ) \,	\coco{U_{\vc k }(\vc x ')}
	& = \delta^{(3)}(\vc x  -\vc x '),
		\\
	% revised 2016 MAR 23
	\label{eq:modenormbasis_k}
	\int_{\Si_\ta}\!\!\!\! \xd^3 x  \; 
	\sqrt{|g^{(3)} g^{\ta \ta}|_{\ta}}\;
	U_{\vc k }(\vc x ) \, \coco{U_{\vc k '}(\vc x )}
	& = \ti w_{\vc k}(\ta)\,\delta^{(3)}(\vc k  -\vc k ').	
	}
Therein, $\om_{\vc k}(\vc x) >0$ is the eigenvalue/eigenfunction of the operator $\om(\vc x)$ upon action on the basis: $w(\vc x)U_{\vc k}(\vc x) = w_{\vc k}(\vc x) U_{\vc k}(\vc x)$, ditto for $\ti w(\ta)$ with eigenvalues $\ti w_{\vc k}(\ta) > 0$.
We require that the product of these two operators, that is, 
of each product $w_{\vc k}(\vc x)\;\ti w_{\vc k}(\ta)$, is $\vc k$-independent 
and yields:
\bal{% revised 2016 MAR 23
	\label{metric_separation}
	w(\vc x)\;\ti w(\ta)
	= \sqrt{|g^{(3)} g^{\ta \ta}|_{\ta}}.
	}
(In Minkowski spacetime, for interval regions bounded by two equal-time hypersurfaces, $\smash{U_{\vc k}(\vc x)}$ corresponds to the set of plane wave modes $\eu^{\im \vc k \,\vc x}\,(2\pi)^{-3}/(2E_{\vc k})$, and $w_{\vc k}(\vc x) = 1/\ti w_{\vc k}(t) = (2\pi)^3(2E_{\vc k})^2$.)
Equation \eqref{eq:modenormbasis_k} shows that the relation \eqref{metric_separation} is actually rather natural: Since we integrate only over $\vc x$, the dependences on $\ta$ and $\vc k$ may remain. 
$w\lvc k(\vc x)$ later serves as a weight function when integrating over $\vc x$ on surfaces $\Si_\ta$ of constant $\ta$, and $\ti w_{\vc k}(\ta)$ serves as a weight function for a Wronskian, which is defined below.
Using these modes, a field configuration $\vph(\vc x )$ on $\Si_\ta$ has the following decomposition (for discrete values of the parameters $\vc k$, the integral is to be replaced by the corresponding sum):
% We need to introduce $w_{\vc k}(\vc x)$ in the equations below,
% because else we cannot even define the usual Fourier trafo in QFT,
% that is: the first two equations in III.C of \cite{Oe:timelike},
%
%
\bal{% revised 2016 MAR 23
	\label{eq:modeinterval}
	\vph(\vc x ) & = \int\!\! \xd^3 k  \; 
	\vph_{\vc k } \, U_{\vc k }(\vc x ),
		&
	% revised 2016 MAR 23
	\vph_{\vc k} & = \int\!\! \xd^3 x\; w_{\vc k}(\vc x)\,
	\vph(\vc x)\, \coco{U_{\vc k}(\vc x)}.
	}
(In Minkowski spacetime this corresponds to spatial Fourier transformation, see Sec. \ref{sec:intervalMindS}.) 
The reality of the field then imposes the reflection property $\vph_{-\vc k} = \coco{\vph_{\vc k}}$, which induces the same property for $w$ and $\ti w$.
We require that the chosen foliation allows for separation of variables 
with respect to the foliation parameter $\ta$ 
in the linear Klein-Gordon equation~\eqref{eq:KG}.
In that case, the modes $U\lvc k(\vc x)$ are essentially eigenfunctions
of the Laplace-Beltrami operator $\De\lvc x$ on the leaves.
Then, by setting $\phi\lvc k(\ta,\vc x) = X\lvc k(\ta)\,U\lvc k(\vc x)$
with fixed $\vc k$ in \eqref{eq:KG}, we obtain an ordinary 
differential equation of second order in $\ta$.
It has two linear independent solutions which we denote
by $X\htx a \lvc k(\ta)$ and $X\htx b \lvc k(\ta)$.
(The superscripts a,b are just labels, not indices.)
They induce the associated operators $X\htx{a}(\ta)$ and $X\htx{b}(\ta)$,
which act on the modes $U\lvc k$ through their eigenvalues 
$X\htx{a}\lvc k(\ta)$ and $X\htx{b}\lvc k(\ta)$:
\bal{% clear
	\label{X_action_eigenvalues_642}
	X\htx a(\ta)\,U_{\vc k}(\vc x) 
	& = X\htx a_{\vc k}(\ta)\,U_{\vc k}(\vc x),
	&
	X\htx b(\ta)\,U_{\vc k}(\vc x) 
	& = X\htx b_{\vc k}(\ta)\,U_{\vc k}(\vc x).
	}
Because of the linearity of the Klein-Gordon equation,
we now assume that any solution of it can be written as follows,
which generalizes Eq.~(16) and the second equation below (37) 
in \cite{Oe:KGtl},	and Eqs.~(8) and (73) in \cite{CoOe:smatrix}:
\bal{% clear
	\label{eq:clsol}
	\phi(\ta, \vc x ) 
	= \biglrr{ X\htx a(\ta) Y\htx a}(\vc x ) 
	+ \biglrr{ X\htx b(\ta) Y\htx b}(\vc x ).
	}
Therein, $X\htx a(\ta),X\htx b(\ta)$ are understood as linear operators 
from the space of real-valued data $Y\htx a(\vc x), Y\htx b(\vc x)$ 
to Klein-Gordon solutions $\phi(\ta,\vc x)$. 
In particular, $X\htx{a}(\ta)$ and $X\htx{b}(\ta)$ acts as operators 
on a mode decomposition of $Y\htx{a}(\vc x)$ and $Y\htx{b}(\vc x)$, 
respectively, as in \eqref{eq:modeinterval}.
%(For Minkowski time-interval regions we have $X\htx a_{\vc k}(t) = \cos(E_{\vc k}t)$ and $X\htx b_{\vc k}(t) = \sin(E_{\vc k}t)$, which are invariant under a sign change of $\vc k$).
Let us call $Y\htx{a}(\vc x)$ and $Y\htx{b}(\vc x)$ reduced data, 
since their values are not actual data on some hypersurface
(values of Klein-Gordon solutions at some fixed $\ta = T$). 
The actual data are generated from the reduced ones only through the action 
of the classical solution operators $X\htx{a}(\ta)$ and $X\htx{b}(\ta)$.

By linearity of the Klein-Gordon operator $(\si_{00}\Box+m^2)$, we can view the functions $\left( X\htx a(\ta) Y\htx a\right)(\vc x )$ and $\left( X\htx b(\ta) Y\htx b\right)(\vc x )$ as two independent solutions of the Klein-Gordon equation \eqref{eq:KG}.
Since the Klein-Gordon equation is of second order, we can assume that the (a priori complex) functions $\smash{X\htx a_{\vc k}(\ta), X\htx b_{\vc k}(\ta)}$ are linear independent for all $\vc k$.
The fact that the field is real requires that 
$X\htx{a}_{-\vc k}(\ta) = \coco{X_{\vc k}\!\htx{a}(\ta)}$ and 
$X\htx{b}_{-\vc k}(\ta) = \coco{X_{\vc k}\!\htx{b}(\ta)}$.
Moreover, we can actually choose the $X\htx a_{\vc k}(\ta)$ and $\smash{X\htx b_{\vc k}(\ta)}$ to be real functions, which makes
$\smash{X\htx{a}(\ta)}$ and $\smash{X\htx{b}(\ta)}$ into real operators.
This can be justified by considering their real and imaginary parts, and knowing that only two of these four functions are linear independent. 
Being real functions makes these eigenvalues independent of the momentum's overall sign, giving the reflection properties
\bal{% clear
	\label{eq:Xab_k_reflect}
	X\htx{a}_{-\vc k}(\ta) & = X\htx{a}_{\vc k}(\ta),
		&
	X\htx{b}_{-\vc k}(\ta) & = X\htx{b}_{\vc k}(\ta).
	}
In the following we shall also assume that $X\htx a(\ta)$ and $X\htx b(\ta)$ commute. 
%They commute, because complex multiplication of functions is commutative.
In the case of non-commuting operators, the computations become more involved, as can be seen in Appendix A of \cite{CoOe:unit}.
Klein-Gordon solutions on an interval region thus write as the following expansion (which we call "real" expansion, since $\smash{X\htx{a}_{\vc k}(\ta)}$ and
$\smash{X\htx{b}_{\vc k}(\ta)}$ are real, but recall that the $U_{\vc k}(\vc x)$ are complex):
\bal{% clear
	\label{eq:modeinterval_sol}
	\phi(\ta,\vc x ) 
	& = \int\!\! \xd^3 k  \, \biiglrr{
		\ph\htx a_{\vc k} \,X\htx a_{\vc k}(\ta)\, U_{\vc {k}}(\vc {x})
		+ \ph\htx b_{\vc k} \,X\htx b_{\vc k}(\ta)\, U_{\vc {k}}(\vc {x})
		}.
	}
The solution's expansion coefficients $(\ph\htx a_{\vc k},\ph\htx b_{\vc k})$ are recovered from initial data $\biglrr{\ph(T,\vc x),(\del_\ta\ph)(T,\vc x) }$ on a hypersurface $\Si_{\ta=T}$ by
\renewcommand\arraystretch{1.5}% make more space in matrices
\bal{% revised 2016 MAR 24
	\label{recover_momrep_interval}
	\begin{pmatrix}
		\ph\htx a\lvc k \\ \ph\htx b\lvc k
	\end{pmatrix}
	& = \int_{\Si_T}\!\! \xd^3 x \;
		w_{\vc k}(\vc x)\, \coco{U\lvc k (\vc x)}\;
		\fracw1{\mc W(T)}
		\begin{pmatrix}
		\;\;(\del_\ta X\htx b)(T) & - X\htx b(T)
			\\
		-(\del_\ta X\htx a)(T) & \;\;X\htx a(T)
		\end{pmatrix}
		\begin{pmatrix}
			\ph(T,\vc x) \\ (\del_\ta\ph)(T,\vc x)
		\end{pmatrix},
	}
\renewcommand\arraystretch{1.0}%
%
%This can be checked by expanding $\ph(T,\vc x)$ and $(\del_\ta\ph)(T,\vc x)$ in \eqref{recover_momrep_interval} using \eqref{eq:modeinterval}, that is, treating them as field configurations  on $\Si_T$ instead as solutions, and then plugging \eqref{recover_momrep_interval} into \eqref{eq:modeinterval_sol} respectively into $\del_\ta$\eqref{eq:modeinterval_sol}.
%Evaluating the resulting expression at $\ta = T$ after using \eqref{eq:modenormbasis} recovers $\ph(T,\vc x)$ respectively $(\del_\ta\ph)(T,\vc x)$, which validates \eqref{recover_momrep_interval}.
%Note also that \eqref{recover_momrep_interval} is well defined, since the following Wronskians never vanish (making them invertible) due to the linear independence of $X\htx a$ and $X\htx b$:
where $\mc W$ denotes the following Wronskian
\balsplit{% clear
	\label{eq:_Wronski_gen}
	\mc W(\ta)
	& \defeq \mcW01\ta{}{}
	= \biglrr{X\htx a\,\del_\ta X\htx b
				-X\htx b\, \del_\ta X\htx a
				}(\ta),
		\\
	% clear
	\mc W_{\vc k}(\ta)
	& \defeq \mc W_{\vc k}^{01}(\ta,\ta)
	= \biglrr{X\htx a_{\vc k}\,\del_\ta X\htx b_{\vc k}
		-X\htx b_{\vc k}\del_\ta X\htx a_{\vc k}}(\ta).
	}
Notice that these quantities never vanish (making them invertible) due to the linear independence of $X\htx a$ and $X\htx b$.
Therein we denote a type of generalized Wronskian by 
%(with $j,k \in \mathds N_0$ the powers of the derivatives, and like the $X\htx{a,b}$, these $\mc W$ are understood as operators acting through eigenvalues, see \eqref{X_action_eigenvalues_642}, and we assume that \emph{all of} these operators are invertible):
%
\bal{% clear
\label{eq:wron-components}
	{\mc W}^{jk}(\ta_1,\ta_2) 
	\defeq \biglrr{\del_\ta^j X\htx a}(\ta_1)\cdot
		\biglrr{\del_\ta^k X\htx b}(\ta_2)\,
		-\,\biglrr{\del_\ta^k X\htx a}(\ta_2)\cdot
		\biglrr{\del_\ta^j X\htx b}(\ta_1).
	}
These $\mc W$ are understood as operators acting through eigenvalues, see \eqref{X_action_eigenvalues_642}, and we formally treat \emph{all of} these operators as invertible.
While \eqref{recover_momrep_interval} recovers a solution from initial Cauchy data,
(that is, the field's value and derivative on a hypersurface of constant $\ta=T$,
 which is not necessarily a Cauchy surface), 
we now want to recover a solution from purely Dirichlet boundary conditions 
(field's value on the boundary hypersurfaces $\Si_1$ and $\Si_2$ 
 of the interval region).
The solution \eqref{eq:clsol} can be expressed in terms of such boundary field configurations $\vph_1(\vc x ) := \phi(\ta_1, \vc x )$ and $\vph_2(\vc x ) := \phi(\ta_2, \vc x )$ as
\bal{		% revised 2016 MAR 24
	\label{eq:_class_sol_boundary_config_intval}
	\phi(\ta,\vc x)
	= \Bigl( \tfracw{\mcW00\ta{}2}{\mcW00\ta12} \vph_1 \Bigr) (\vc x)
		+\Bigl( \tfracw{\mcW00\ta1{}}{\mcW00\ta12} \vph_2 \Bigr) (\vc x).
	}
For Minkowski spacetime, this formula is given for example in Eq.(13-25) of Hatfield's QFT book \cite{Hat}. 
We remark that the only goal of \eqref{eq:_class_sol_boundary_config_intval} is to express the action of the classical solution $\ph$ in terms of its boundary configuration $\vph$.
While the solution $\ph$ is fixed by $\vph$ only up to adding a solution vanishing on the boundary, the action is fixed by $\vph$ completely, as we have discussed below expression \eqref{eq:_ph_vph}.
Further, the operator $\mcW00\ta12$ is not strictly invertible, since  its eigenvalues $\mc W^{00}\lvc k (\ta_1,\ta_2) := X\htx a\lvc k(\ta_1)X\htx b\lvc k(\ta_2)-X\htx a\lvc k(\ta_2)X\htx b\lvc k(\ta_1)$ might vanish for some momenta $\vc k$. 
Therefore we regard the inverse of $\mc W^{00}(\ta_1,\ta_2)$ as an intermediate means, which does not appear in the definitions of quantum states and the final formulas for the amplitudes.
The solution \eqref{eq:_class_sol_boundary_config_intval} allows to evaluate the action \eqref{eq:actintreg}, resulting in
\bal{	% revised 2016 MAR 24
	S^0_{[\ta_1,\ta_2]}(\phi)
	=  \frac{1}{2} \int\!\! \xd^3 x  \, 
		\begin{pmatrix} \vph_1 & \vph_2 \end{pmatrix} 
		W_{[\ta_1,\ta_2]}  
		\begin{pmatrix} \vph_1 \\ \vph_2 \end{pmatrix}.
	\label{eq:actev}
}
Therein, $W_{[\ta_1,\ta_2]}$ is a $(2,2)$-matrix with operator-valued elements $W_{[\ta_1,\ta_2]}^{ij}$ with $i,j\in\{1,2\}$ given by
\begin{align}
% revised 2016 MAR 24
W_{[\ta_1,\ta_2]}^{11} 
=& -\si \sqrt{|g^{(3)}g^{\ta \ta}|_{\ta\!=\!\ta_1}} \,
	\tfrac{\mcW10\ta12}{\mcW00\ta12},
	& 
W_{[\ta_1,\ta_2]}^{12} 
=& -\si \sqrt{|g^{(3)} g^{\ta \ta}|_{\ta\!=\!\ta_1}} \,
	\tfrac{\mcW01\ta11}{\mcW00\ta12},
	\notag
	\\
\label{eq:W2}
W_{[\ta_1,\ta_2]}^{21} 
=&+\si\sqrt{|g^{(3)} g^{\ta \ta}|_{\ta\!=\!\ta_2}} \,
	\tfrac{\mcW10\ta22}{\mcW00\ta12},
	&
W_{[\ta_1,\ta_2]}^{22} 
=& +\si \sqrt{|g^{(3)} g^{\ta \ta}|_{\ta\!=\!\ta_2}}\,
	\tfrac{\mcW01\ta12}{\mcW00\ta12}.
\end{align}
In order to show the symmetry of this matrix, let us consider the symplectic structure on the space of smooth Klein-Gordon solutions on $M\lintval\ta12=[\ta_1,\ta_2] \times I^{(3)}$. 
It is given by

\bal{% revised 2016 MAR 24
% negative sign due to our backward orientation of \Si_\ta
\label{eq:sympl}
\omega(\xi, \ze) 
= -\frac{\si}{2}\int_{\Sigma_\ta}\!\!\!\! \xd^3 x   \, 
\sqrt{|g^{(3)} g^{\ta \ta}|} 
\left(\xi \, \del_{\ta} \ze - \ze \, \del_{\ta} \xi \right).
}
The leaf $\Si_\ta$ is canonically oriented in negative $\ta$-direction (that is, backwards). 
In case of the leaves being spacelike, this is just the standard symplectic form.
Using mode decomposition \eqref{eq:modeinterval_sol} and orthogonality \eqref{eq:modenormbasis_k}, the symplectic structure evaluates to

\bal{\om \biglrr{\xi,\ze}
	% revised 2016 MAR 24
	\label{SKG_structures_KleiGo_33}	
	& = -\frac{\si}2\,
	\int\!\!\xd^3 k\, \biiglrr{ \xi\htx a_{\vc k}\,\ze\htx b_{-\vc k}\,
									- \xi\htx b_{\vc k}\,{\ze\htx a_{-\vc k}}  
									}\,
	\ti w_{\vc k}(\ta) \mc W\lvc k (\ta).
	}
The symplectic structure \eqref{eq:sympl} is independent of the leaf $\Sigma_\ta$ of the foliation chosen to integrate over (with $\ta\in[\ta_1,\ta_2]$, see for example \cite{Woo:geomquant}). 
Therefore, the weighted Wronskian $\ti w_{\vc k}(\ta) \mc W\lvc k (\ta)$ must be independent of $\ta$.
This implies that the operators $\sqrt{|g^{(3)} g^{\ta \ta}|_{\ta}}\, \mcW01\ta{}{} =-\sqrt{|g^{(3)} g^{\ta \ta}|_{\ta}}\, \mcW10\ta{}{}$ are independent of $\ta$ as well.
This causes the operator equality $W_{[\ta_1,\ta_2]}^{12}=W_{[\ta_1,\ta_2]}^{21}$, which shows that the matrix $W_{[\ta_1,\ta_2]}$ is symmetric.
Note also that due to their definition via multiplication with eigenvalues/functions, the operators $X\htx{a}(\ta)$, $X\htx{b}(\ta)$ 
and all related operators such as $\mc W^{jk}(\ta_1,\ta_2)$ and $W^{jk}\lintval\ta12$ etc., are symmetric with respect to the inner product $\inpro{\vph}{\ki}=\tfrac12\,\int\!\xd^3x\;\vph(\vc x)\,\ki(\vc x)$.
% We frequently use this symmetry property in our calculations.
%
% ==============================================================
%
\subsection{Hypercylinder regions: rods and tubes}
\label{sec:hypercylinder}
\noindent
In order to define two more types of regions, we again introduce a foliation of the spacetime, defined by a smooth coordinate system $(t,r,\upte, \upvph)$. Therein, $t \in \reals$ is now a time variable and $r \in [0,\infty)$ is a radial coordinate. 
$\upte \in [ 0, \pi ]$ and $\upvph \in [0, 2 \pi)$ are angular coordinates, for which we use the collective notation $\Omega\defeq(\upte,\upvph)$ and $\xd\Om = \xd\upte\, \xd\upvph$. 
% The sine in Minkowski case is not included in $\xd \Om$,
% it comes from the term $\sqrt{g^{(3)}}$.
The leaves of this foliation are the hypersurfaces of constant $t$. 
Our new regions are defined in terms of hypercylinders $\Si_r$, which are the hypersurfaces of constant radius $r$, that is: $\Si_r = I^{(t)} \times \sphere[2]_r$, wherein $I^{(t)}\subseteq\reals$ represents all of time.
The hypercylinders $\Si_r$ are canonically oriented in direction of negative $r$, that is, inwards.
Here, we require the metric to be block diagonal with respect to the radial coordinate, that is: $0=g^{tr}=g^{r\upte}=g^{r\upvph}$.
Note that this is fulfilled e.g.~by Anti de Sitter and several
black hole metrics, including Kerr-Newman.
% hypercylinders do not provide a foliation of the whole spacetime,
% since they degenerate at the time axis r=0

As a first type of regions, we can define the so called tube regions: 
These regions are bounded by two concentric hypercylinders of different radii $R_1$ and $R_2$.
Hence they are a radial analogue of the interval regions defined above. 
The quantities associated to tube regions $M_{[R_1,R_2]} \defeq I^{(t)} \times [R_1,R_2] \times \sphere[2]$ are labeled by $[R_1,R_2]$
and the boundary writes $\del M\lintval R12 = \Si_{R_1} \cup \Sibar_{R_2}$.

The second type of region $M_R = I^{(t)} \times [0,R] \times \sphere[2]$ is called rod region and is bounded by \emph{only one} hypercylinder, namely $\del M_R=\Sibar_R$. 
We will use the label $R$ for the quantities associated to the rod region $M_R$. 
Notice the connectedness of the boundary of the spacetime region in which the dynamics of the scalar field is considered. 
We assume that we can cover the whole spacetime with a rod region by sufficiently increasing $R$.
%(With a tube region this is never possible: decreasing $R_1$ towards 0 finally turns it into a rod region.)

The free action of a Klein-Gordon field in these regions takes the following form, wherein $g^{(3)}$ is the induced metric on the hypercylinders of the respective fixed radii,

\bal{	\label{eq:actintreg2}
	% clear
	S^0_{[R_1, R_2]} (\phi) 
	&= - \frac{1}{2}\! \int\!\! \xd t \, \xd \Omega\, \biiiglrr{\!\! 
		\sqrt{|g^{(3)}g^{rr}|_{r=R_2}} \,
		\biglrr{\phi\, \del_{r} \phi} (t,R_2, \Omega) 
		- \!\sqrt{|g^{(3)}g^{rr}|_{r=R_1}} \,
		\biglrr{\phi\, \del_{r} \phi} (t,R_1, \Omega)\!
		},
		\\
	\label{eq:actintreg3}
	% clear
	S^0_{R} (\phi)
	&= - \frac{1}{2} \int\!\! \xd t \, \xd \Omega \; 
	\sqrt{|g^{(3)}g^{rr}|_{r=R}}\;
	\, \biglrr{\phi\, \del_{r} \phi} (t,R, \Omega).
	}
Again we introduce mode decompositions for the Klein-Gordon solutions and for the boundary field configurations. 
Since here the foliation involves the sphere $\sphere[2]$, the corresponding momenta are now discrete, and we shall denote them like the angular momentum numbers in Minkowski spacetime simply by $l$ and $m_l$ (the subscript $l$ distinguishes $m_l$ from the field mass $m$,  and we assume without loss of generality that $l\in\mathds N_0$ and $m_l\in\{-l,-l\!+\!1,\ldots,+l\}$).
Since $t$ usually takes values on the whole real line, we assume the corresponding momentum $\om$ to be continuous.
In analogy to \eqref{eq:modenormbasis}, we assume a set of complex modes $\{U_{\om l m_l\!}(t,\Om)\}$ fulfilling the reflection property $U_{\!-\!\om, l, \!-\! m_l}(t,\Om)=\coco{U_{\om l m_l\!}(t,\Om)}$, which forms a complete orthonormal basis in the space of field configurations on the hypercylinders $\Si_r$, and also in momentum space:
\bal{\label{eq:modenormbasis_cyl}
	% revised 2016 MAR 25
	\int\!\! \xd\om \sum_{l,m_l} w_{\om l m_l\!}(t,\Om)\,
	U_{\om l m_l\!}(t,\Om) \,
	\coco{U_{\om l m_l\!}(t',\Om')}
	& = \delta(t\!-\! t')\,\delta^{(2)}(\Om,\Om')
		\\
	% revised 2016 MAR 25
	\label{eq:modenormbasis_cyl_omlml}
	\int_{\Si_R}\!\!\!\! \xd t\,\xd\Om \sqrt{|g^{(3)} g^{rr}|_{R}}\;
	U_{\om l m_l\!}(t,\Om) \,
	\coco{U_{\om' l' m'_l\!}(t,\Om)}
	& = \ti w_{\om l m_l\!}(R)\,\delta(\om\!-\!\om')\,\de_{ll'}\,\de_{m_l m'_l}.	
	}
(For Minkowski hypercylinders, $U_{\om l m_l\!}(t,\Om)$ are the modes $\eu^{\!-\!\im \om t}\,Y^{m_l}_l(\Om)$, wherein $Y^{m_l}_l(\Om)$ denotes the spherical harmonics on $\sphere[2]$.)
Again we require the product $w\ti w$ to yield the metric root:
\bal{		% clear
	\label{metric_separation_r}
	w(t,\Om)\, \ti w(R)
	= \sqrt{|g^{(3)} g^{rr}|_{R}}.
	}
Using these modes, a field configuration $\vph(t,\Om)$ on $\Si_r$ has the following decomposition:

\bal{		% clear
	\label{eq:modeshyp}
	\vph(t,\Om) & = \int\!\!\xd\om\sum_{l,m_l}
	\vph_{\om l m_l\!}\, U_{\om l m_l\!}(t,\Om), \quad
		&
	\vph_{\om l m_l\!} & = \int\!\!\xd t\,\xd\Om\;
	w_{\om l m_l\!}(t,\Om)\;
	\vph(t,\Om)\; \coco{U_{\om l m_l\!}(t,\Om)}\;.
	}
The reality of the field imposes the reflection properties $\vph_{\!-\!\om,l,\!-\! m_l} = \coco{\vph_{\om l m_l\!}}$, ditto for $w$ and $\ti w$.
(In Minkowski spacetime this corresponds to spatial spherical harmonic decomposition plus temporal Fourier transformation.)
As for interval regions, we require that we can apply separation of variables 
and that any Klein-Gordon solution on a tube region can be written as
\bal{% clear
	\phi(t,r,\Om) = 
	\biglrr{ X\htx a(r) Y\htx a}(t,\Om)
	+ \biglrr{ X\htx b(r) Y\htx b}(t,\Om),
	\label{eq:clsol_tube}
	}
whereas a Klein-Gordon solution on a rod region can be written as
\bal{% clear
	\phi(t,r,\Om) = \biglrr{ X\htx a(r) Y\htx a}(t,\Om).
	\label{eq:clsol_rod}
}
Expression \eqref{eq:clsol_rod} needs a comment. 
In the cases studied so far, i.e.~a scalar theory in Minkowski space \cite{Oe:KGtl,CoOe:smatrix,CoOe:letter}, in de Sitter space \cite{Co:dS,Co:letter} and in 2d Euclidean space \cite{CoOe:2d}, the Klein-Gordon equation expressed in spherical (polar in 2d) coordinates reduces to a certain Bessel equation, with two independent solutions provided by the spherical Bessel functions of the first and second kind (Neumann functions) respectively. 
These functions have different behavior at the origin: The former is regular, while the latter diverges at the origin.
Since the rod hypercylinder region (disk region in the 2d Euclidean theory) contains the origin, only the spherical Bessel functions (first kind) are admissible to obtain a smooth solution of the Klein-Gordon equation on this region, while the Neumann functions are not. 
The same happens for Anti de Sitter (AdS) spacetimes with different types of hypergeometric functions taking the roles of spherical Bessel and Neumann functions \cite{dohse:_class_AdS}, and for Rindler spacetime~\cite{CoRa:rindler}.
We are assuming a similar situation here, where $X\htx a$ represents the regular solution to the radial part of the Klein-Gordon equation, while $X\htx b$ represents the diverging solution.%
\footnote{This will be the case for spaces conformal to (a portion of) Minkowski spacetime. However, we can also consider a more general situation where both $X\htx a$ and $X\htx b$ result to be well defined in the whole rod region. In this case the solution (\ref{eq:clsol_rod}) can be expressed in terms of a linear combination of $X\htx a$ and $X\htx b$. We shall not elaborate on this aspect here.}
Klein-Gordon solutions on a tube region can then be written as an 
expansion like \eqref{eq:modeinterval_sol}:
\bal{\label{eq:modetub}
	% clear
	\phi(t,r,\Om) 
	& = \int\!\!\xd\om\sum_{l,m_l}
	\biglrr{\ph\htx a_{\om l m_l\!} \,X\htx a_{\om l m_l\!}(r)\, 
				U_{\om l m_l\!}(t,\Om)
			+ \ph\htx b_{\om l m_l\!} \,X\htx b_{\om l m_l\!}(r)\, 
			U_{\om l m_l\!}(t,\Om)
			}.
	}
The expansion coefficients $(\ph\htx a_{\om l m_l\!},\ph\htx b_{\om l m_l\!})$ of the solution can be recovered from initial data $\biglrr{\ph(t,R,\Om),(\del_r\ph)(t,R,\Om) }$ on a hypercylinder $\Si_{R}$ by \eqref{recover_momrep_interval}:

\renewcommand\arraystretch{1.3}% make more space in matrices
\bal{	% revised 2016 MAR 25
	\label{recover_momrep_tube}
	\begin{pmatrix}
	\ph\htx a_{\om l m_l\!} \\
	\ph\htx b_{\om l m_l\!} 
	\end{pmatrix}
	& = \int_{\Si_R}\!\!\!\!\xd t\,\xd\Om\; w_{\om l m_l\!}(t,\Om)\, 
		\coco{U_{\om l m_l\!} (t,\Om)}\,\frac1{\mc W(R)}
	\begin{pmatrix}
	(\del_r X\htx b)(R) & -X\htx b(R) \\
 	- (\del_r X\htx a)(R) & X\htx a(R)
	\end{pmatrix}
	\begin{pmatrix}
	\ph(t,R,\Om) \\
	(\del_r\ph)(t,R,\Om)
	\end{pmatrix}.
	}
\renewcommand\arraystretch{1.0}%
Klein-Gordon solutions on a rod can also be expanded as 
%a "real" expansion:
%
\bal{\label{eq:moderod}
	% clear
	\phi(t,r,\Om) 
	& = \int\!\!\xd\om\sum_{l,m_l}
	\ph\htx a_{\om l m_l\!} \,X\htx a_{\om l m_l\!}(r)\, U_{\om l m_l\!}(t,\Om),
	}
and the expansion coefficient $\ph\htx a_{\om l m_l\!}$ of the solution can be recovered from Dirichlet boundary data $\ph(t,R,\Om)$ on a hypercylinder $\Si_{R}$ by
\balsplit{ % revised 2016 MAR 25
	\label{recover_momrep_rod}
	\ph\htx a_{\om l m_l\!}
	& = \int_{\Si_R}\!\!\xd t\,\xd\Om\; w_{\om l m_l\!}(t,\Om)\; 
			\coco{U_{\om l m_l\!} (t,\Om)}\;
			\left( X\htx a(R)\right)^{-1}\,\ph(t,R,\Om)\;.
	}
For the tube hypercylinder regions $M_{[R_1,R_2]}$, a solution expressed in terms of its boundary configurations reads like \eqref{eq:actev}:
\bal{	% clear
	\label{eq:clsolhyp_tub}
	\phi(r, t, \Omega)
	= \biigl(\! \tfracw{\mc W^{00}(r,R_2)}{\mcW00R12}\,
		\vph_1\biigr) (t, \Omega) 
	+ \biigl(\! \tfracw{\mc W^{00}(R_1,r)}{\mcW00R12} \,
		\vph_2\biigr) (t, \Omega), 
	}
while for a solution on a rod region $M_R$ we have instead
\bal{% clear from {eq:moderod} and {recover_momrep_rod}
	\label{eq:clsolhyp_rod}
	\phi (t,r, \Omega) 
	= \left(\! \tfracw{X\htx a(r)}{X\htx a (R)}\,
	 \vph \right) (t, \Omega).
	}
Expression \eqref{eq:clsolhyp_tub} and \eqref{eq:clsolhyp_rod} allow to express the action of the classical solution $\ph$ in terms of its boundary configuration $\vph$.
The operator $X\htx a(R)$ is not strictly invertible, since  its eigenvalues $X\htx a_{\om l m_l\!}(R)$ might vanish for some momenta $(\om,l,m_l)$.
Therefore we also regard the inverse of $X\htx a(r)$ as an intermediate means, which does not appear in the definitions of quantum states and the final formulas for the amplitudes.
We can now write the action for the tube region as in \eqref{eq:actev} for the interval region,
\bal{
	% clear
	S^0_{[R_1, R_2]}(\phi)
	=  \frac{1}{2} \int \xd t \, \xd \Omega \, 
	\begin{pmatrix}\vph_1 & \vph_2 \end{pmatrix} W_{[R_1,R_2]}  
	\begin{pmatrix} \vph_1 \\ \vph_2 \end{pmatrix},
	\label{eq:actevhyp}
}
with the matrix elements derived from \eqref{eq:W2}:
\begin{align}
	% clear
	W_{[R_1,R_2]}^{11} 
	=& + \sqrt{|g^{(3)}g^{rr}|_{r\!=\! R_1}} \,
		\tfracw{\mcW10R12}{\mcW00R12},
		& 
	W_{[R_1,R_2]}^{12} 
	=& + \sqrt{|g^{(3)} g^{rr}|_{r\!=\! R_1}} \,
		\tfracw{\mcW01R11}{\mcW00R12},
		\notag
		\\
	% clear
	\label{eq:W3}
	W_{[R_1,R_2]}^{21} 
	=&-\sqrt{|g^{(3)} g^{rr}|_{r\!=\! R_2}} \,
		\tfracw{\mcW10R22}{\mcW00R12},
		&
	W_{[R_1,R_2]}^{22} 
	=& -\sqrt{|g^{(3)} g^{rr}|_{r\!=\! R_2}}\,
		\tfracw{\mcW01R12}{\mcW00R12}.
	\end{align}
The free action of the field in the rod region of radius $R$ is
\bal{\label{freeaction_rod_boundary}
	% clear
	S^0_{R}(\phi) & = \frac{1}{2} \int \xd t \, \xd \Omega \,
  \; \vph_R(t, \Omega) \left( W_R\, \vph_R \right) (t, \Omega),
		\\
	% revised 2016 MAR 25
	W_R & = -\sqrt{|g^{(3)}g^{rr}|_{r\!=\! R}}\;
	\tfracw{\left(\del_r X\htx a\right)(R)}{X\htx a(R)}.
	}
By the same argument as for the interval regions, the weighted Wronskian $\ti w(r) \mc W_{\om l m_l\!} (r)$ is independent of $r$, and thus the operators $\sqrt{|g^{(3)} g^{rr}|_{R}}\; \mcW01R{}{}
 =-\sqrt{|g^{(3)} g^{rr}|_{R}}\; \mcW10R{}{}$ are independent of $R$ as well, causing the operator equality $W_{[R_1,R_2]}^{12}=W_{[R_1,R_2]}^{21}$ and making $W_{[R_1,R_2]}$ a symmetric matrix.
%
% ==============================================================
%
\subsection{Well-posedness of the initial value problem}
\label{sec:wellposed}
\noindent
In Sections~\ref{sec:interval} and \ref{sec:hypercylinder} we consider
classical solutions which are determined by initial data on hypersurfaces
that are either spacelike or timelike.
Well-posedness means, that a solution to the initial value problem exists,
and is both unique and stable (depends continuously on the initial conditions).
For hyperbolic equations like the Klein-Gordon equation~\eqref{eq:KG},
it is known that %they 
initial value problems 
are well-posed for compactly supported initial data
on Cauchy surfaces. However, they are typically not well-posed 
for initial data on timelike hypersurfaces. 
Nevertheless, a well-defined solution of the Klein-Gordon equation may exist for specific boundary data. In particular in the examples treated so far in literature, namely the Klein-Gordon theory in 
%We cannot offer a general solution to this open problem.
%Rather, we show how this problem is circumvented in several examples
%(
Minkowski, Rindler, de Sitter and Anti de Sitter spacetimes, the 
%).
%We assume that the 
properties of the functions $X\htx a_{\vc k}(\ta)$
and $X\htx b_{\vc k}(\ta)$ were assumed to define bounded solutions in the spacetime regions considered. Following the same line of reasoning we make the same assumption here.
%which make this possible in the examples,
%hold for many more spacetimes. 
For spacetimes where this does not apply, 
the methods presented in this article will cease to work.

We first consider the interval regions of Section~\ref{sec:interval}
for the case of timelike boundaries.
Denoting the chosen time coordinate by $y^0 = t$ 
and the spatial coordinates by $y^1, y^2, y^3$,
in the notation of Section~\ref{sec:interval} we then have 
$\ta = y^1$ and $\vc x = (t, y^2, y^3)$.
For Minkowski spacetime, this case has been treated 
in Section IV of \cite{Oe:KGtl}.
A classical Klein-Gordon solution $\phi$ is determined by the initial
data $\biglrr{\ph(T,\vc x),(\del_\ta\ph)(T,\vc x) }$
on a constant-$y^1$ hypersurface $\Si_{T}$
through the linear equation~\eqref{recover_momrep_interval}
together with the linear mode expansion~\eqref{eq:modeinterval_sol}.
Assuming that the integrals therein converge, existence and uniqueness
are satisfied, and we still need to ensure stability.

The problem with stability here arises due to the fact, 
that for some momenta the functions $\smash{X\htx a\lvc k(y^1)}$ and 
$\smash{X\htx b\lvc k(y^1)}$ are not bounded on all of spacetime.
For example, in Minkowski spacetime, there appear evanescent modes 
with $E^2 < m^2$ exhibiting exponential behaviour like 
$\smash{\phi(y) \sim \eu^{\ka y^1}}$ for some $\ka \in \reals$.
We might now feed Eq.~\eqref{recover_momrep_interval}
with special initial data $\biglrr{\ph(T,\vc x),(\del_\ta\ph)(T,\vc x) }$,
that induces a bounded solution (does not induce evanescent modes 
in the solution). Let us refer to such data as 'finetuned'. 
For the above Minkowski example, such initial data can be easily
generated by evaluating Eq.~\eqref{eq:modeinterval_sol} 
at fixed $\ta = T$ with the integration restricted to $\vc k$
with $\vc k^2 > m^2$. However, even a small variation of such 
finetuned data might induce a solution that also contains modes 
which are not bounded on all of spacetime, 
and this renders the solution unstable.

However we notice that this problem will not show up if we consider
%This problem can be circumvented by considering the 
classical solutions 
not on the whole spacetime, but only on an interval region of spacetime 
with $y^1 \in [y^1_1, y^1_2]$ and $y^1_1 \leq T \leq y^1_2$.
Assuming that the functions $X\htx a\lvc k(y^1)$ and $X\htx b\lvc k(y^1)$
do not diverge on this interval, the solution is now stable
when considered only on this spacetime region.
For spacetime regions without curvature singularities,
%we see no reason why bounded initial data should
%produce an infinite Klein-Gordon field, and therefore 
we assume
that no divergencies occur for solutions defined by bounded initial data in cartesian coordinates
as used in Section~\ref{sec:interval}. However, we know 
that divergencies do occur when radial coordinates are used.
For example, on Minkowski spacetime the radial functions
$\smash{X\htx a_{\om l m_l}(r)}$ and $\smash{X\htx b_{\om l m_l}(r)}$ 
are spherical Bessel functions respectively spherical Neumann functions,
the former being regular and the latter divergent at the origin.

In order to see how to deal with this issue, let us progress 
to the hypercylinder regions of Section~\ref{sec:hypercylinder}.
The problem here is quite similar: Now Eq.~\eqref{recover_momrep_tube}
determines the solution $\phi$ through initial data on a hypercylinder $\Si_R$.
Starting anew with finetuned initial data, taken as inducing a solution $\phi$
that contains only regular modes, a small variation of the data
might induce a solution that also contains diverging modes,
rendering the solution unstable again.
Since the divergence should occur at the origin $r=0$,
the stability problem is avoided in a similar way as above:
We consider the classical solution only on the tube region
where $r \in [R_1, R_2]$ with $R_1 \leq R \leq R_2$, on which it is stable.
What makes it possible to avoid the stability problem
for interval and tube regions in Minkowski spacetime, is the crucial fact
that Eqs.~\eqref{recover_momrep_interval} and \eqref{recover_momrep_tube}
establish a one-to-one correspondence between arbitrary bounded initial data
and bounded solutions in the regions.

The situation is different however for the rod regions:
The relevant equation is now \eqref{recover_momrep_rod},
which contains the inverse of the operator $X\htx a (R)$.
Since the function $X\htx a_{\om l m_l} (R)$ vanishes for some
momenta $(\om, l, m_l)$, the classical problem would become ill posed
if we were to admit arbitrary bounded initial data $\phi(t,R,\Om)$
in Eq.~\eqref{recover_momrep_rod}.
Equivalently, Eq.~\eqref{eq:clsolhyp_rod} is not well defined
for arbitrary bounded data $\vph(t,\Om)$.
The reason for this is that bounded solutions on rod regions
cannot generate arbitrary bounded data on the boundary hypercylinder $\Si_R$:
They cannot generate data for modes with precisely those 
momenta $(\om, l, m_l)$ for which $X\htx a_{\om l m_l} (R)$ vanishes.
This difficulty can be avoided by admitting only initial data
which is induced by bounded classical solutions. 
That is, data $\vph(t,\Om) = \phi(t,R,\Om)$ which is
generated by evaluating \eqref{eq:moderod} at fixed radius $R$.
For such data, Eqs.~\eqref{recover_momrep_rod} and \eqref{eq:clsolhyp_rod}
become well defined.
We also remark that, as we point out below \eqref{eq:clsolhyp_rod},
this issue of the classical theory does not affect the quantum results,
because the inverse of the operator $X\htx a_{\om l m_l}(r)$ 
neither appears in the definition of our quantum states nor in 
the amplitudes \eqref{zzz_SKG_free_amplitude_100} for rod regions
(and the occurrence of infinities in intermediate results
is rather common in quantum calculations).
%
% ==================================================================
%
\section{Quantization}
\label{sec:quantum}
\noindent
We adopt the Schr\"odinger-Feynman quantization scheme, namely the quantum states of the field are described in the Schr\"odinger representation \cite{CoCoQu:SF, Hat, Jackiw} by wave functionals on spaces of field configurations, and amplitudes are calculated through a path integral quantization. 
According to the axioms of the General Boundary Formulation (GBF), to each oriented hypersurface $\Sigma$ (hypersurfaces have codimension one) we associate a quantum state space $\cHS_{\Sigma}$ of wave functionals of field configurations on $\Sigma$ (the label S is for Schr\"odinger).
Note that in the GBF sense the disjoint unions of hypersurfaces count again as hypersurfaces, and the state space of such a hypersurface is simply the tensor product of the union's constituent hypersurfaces' state spaces.
One particular class of hypersurfaces consists of the boundaries of spacetime regions (regions have codimension zero), and therefore each region $M$'s boundary $\del M$ has its state space $\cHS_{\del M}$.
As usual, we orient boundaries as pointing outwards of the enclosed regions. 
The inner product of the Hilbert space $\cHS_{\Sigma}$ is formally given by 
\bal{
	% clear
	\label{eq:inpro_schro}
	\inpro{\al\htx S_\Si}{\be\htx S_\Si}
	\defeq \int_{K_\Si}\!\!\!\! \xD \vph \; 
	\coco{\al\htx S_\Si(\vph)} \, \be\htx S_\Si(\vph),
	}
where the integral is over the space $K_\Si$ of field configurations $\vph$ on the hypersurface $\Sigma$. 
As familiar in QFT, the inner product \eqref{eq:inpro_schro} often becomes infinite.
We recall that $\Sibar$ denotes the same hypersurface $\Si$ with opposite orientation. 
It also has its associated state space $\cHS_\Sibar$.
Throughout this article, we treat both state spaces as identified $\smash{\cHS_\Sibar = \cHS_\Si}$, writing $\smash{\psS{\Sibar} \defeq \coco{\psS{\Si}}}$.
(However, field configurations do not depend on the orientation of $\Si$, and thus $K_\Sibar \equiv K_\Si$.)

Standard transition amplitudes are generalized in the GBF by amplitudes associated to spacetime regions $M$, given by linear amplitude maps $\rho\htx S_M : \cHS_{\del M} \rightarrow \complex$ from the region $M$'s boundary state space to the complex numbers.
(We emphasize that the GBF is not some special quantum theory, but rather a framework about how to formulate any specific quantum theory.
Hence the amplitude map $\ro_M$ encoding the dynamics taking place inside $M$ is not fixed from the outset, but depends on the specific quantum theory studied in the GBF framework, which in our case happens to be real Klein-Gordon theory).
Boundary state spaces and amplitudes are required to satisfy a number of consistency axioms \cite{Oe:GBQFT}, some of which are considered later in this section.
The amplitude $\roS{M}$ for a boundary state $\psS{\del M}$ is defined heuristically as
\bal{
	% clear
	\label{eq:amplitude_general}
	\roS{M}(\psS{\del M}) = \int_{K_{\del M}}\!\! \xD \vph \; 
						\psS{\del M}(\vph) \, Z_M(\vph),
	}
wherein $Z_M$ is the field propagator encoding the field dynamics in the spacetime region $M$:
\bal{
	% clear
	\label{eq:fieldprop}
	Z_M(\vph) 
	= \int_{\phi|_{\del M} = \vph} \!\!\!\! \xD \phi \; 
		\eu^{\im S_M(\phi)}.
	}
$S_M(\phi)$ is the action of the field in the region $M$, and the integration is extended over all field configurations $\phi$ (not only classical solutions) matching the boundary configuration $\vph$ on the boundary $\del M$. 
Next we consider the above objects for the types of regions introduced in Section \ref{sec:_classical}.

For interval (and tube) regions $\smash{M_{[\ta_1,\ta_2]}}$, the boundary hypersurface is the union of two disjoint hypersurfaces of constant $\ta$ each (hypercylinder surfaces of constant $r$ each), and hence the boundary state space $\cHS_{\del[\ta_1,\ta_2]} = \cHS_{\ta_1} \otimes \cHS_{\ta_2}$ is the tensor product of the two boundary components' state spaces.
% 
% no dual space needed here, actually the second state 
% which we always use is not a dual state 
% but just the complex conjugate of a usual state,
% which of course can be identified via Riesz with a dual state,
% but there is no real need here to mention dual spaces,
% it probably confuses some readers if we write dual space 
% despite not using dual states
%
A state in this Hilbert space thus writes as $\psS{\Si\ta_1} \otimes \smash{\coco{\psS{\Si\ta_2}}}$, wherein the complex conjugation of the second state is due to the opposite orientation of the second hypersurface (because both are oriented outwards).
The amplitude for this state takes the form
\bal{ % clear
	\label{eq:amplitude_general_interval}
	\roS{[\ta_1,\ta_2]} (\psS{\Si\ta_1} 
				\otimes \coco{\psS{\Si\ta_2}}\,) 
	& = \int_{K_1}\!\!\!\! \xD \vph_1 
		\int_{K_2}\!\!\!\!\xD \vph_2 \;
		\psS{\Si\ta_1}(\vph_1) \, \coco{\psS{\Si\ta_2}(\vph_2)} \;
		Z_{[\ta_1,\ta_2]}(\vph_1, \vph_2),
		\\
	% clear
	\label{eq:propinterval}
	Z_{[\ta_1,\ta_2]}(\vph_1, \vph_2) 
	& = \int_{\begin{matrix}
					\scriptstyle	\phi|_{\Sigma_1}=\vph_1 \\ 
					\scriptstyle \phi|_{\Sigma_2}=\vph_2
					\end{matrix}
					} 
	\xD \phi \; \eu^{\im S_{[\ta_1,\ta_2]}(\phi)}.
	}
For tube regions $M_{[R_1,R_2]}$ the boundary state space is 
$\cHS_{\del [R_1,R_2]} = \cHS_{R_1} \otimes \cHS_{R_2}$, 
and a state in this Hilbert space is $\psS{R_1} \otimes \coco{\psS{R_2}}$.
This state's amplitude writes just as \eqref{eq:amplitude_general_interval} with field propagator \eqref{eq:propinterval}, with $\ta_{1,2}$ replaced in both by $R_{1,2}$.

For a rod region $M_R$, the boundary state space is $\cHS_{\del M_{R}} =  \cHS_{R}$, and a state in this Hilbert space is $\coco{\psS{R}}$ since $\Si_R$ is oriented inwards while $\del M_R =\Sibar_R$ is oriented outwards. The state's amplitude is

\bal{	% clear
	\label{eq:amplitude_general_rod}
	\roS{R} (\coco{\psS{\Si_R}}) 
	= \int\!\! \xD \vph_R \;
	\coco{\psS{\Si_R}(\vph_R)} \, Z_{R}(\vph_R),
	}
and the field propagator of the theory reads (with $S_{R}(\phi)$ the action of the rod region)
\bal{
	% clear
	\label{eq:proprod}
	Z_{R}(\vph_R) = \int_{\phi|_{R} = \vph_R} \!\!\xD \phi \; 
							\eu^{\im S_{R}(\phi)}.
	}
%
%Note that here the amplitude is calculated for a single state living on the boundary hypercylinder (and not for two states as usual).
%This boundary state encodes both incoming and outgoing particles.
%As usual, this amplitude thus relates to the probability of observing the outgoing particles given that the incoming ones are prepared.
In addition to determining amplitudes, the field propagator also propagates quantum states across spacetime regions.
For interval regions with $\ta_1<\ta_2$, we obtain a new state $\ps_{\Si\ta_2}$ on $\Si_{\ta_2}$ by propagating the initial state $\ps_{\Si\ta_1}$ from $\Si_{\ta_1}$ across the region $M\lintval\ta12$:
\bal{
	% clear
	\label{eq:state_propagation_interval}
	\psS{\Si\ta_2}(\vph_2)
	& = \int_{K_1}\!\!\!\! \xD\vph_1\;
	\psS{\Si\ta_1}(\vph_1)\,
	Z_{[\ta_1,\ta_2]}(\vph_1,\vph_2),
		\\
	% inverse of above, consistent with unitarity property
	% revised 2016 MAR 26
	\psS{\Si\ta_1}(\vph_1)
	& = \int_{K_2}\!\!\!\! \xD\vph_2\;
	\psS{\Si\ta_2}(\vph_2)\,
	\coco{Z_{[\ta_1,\ta_2]}(\vph_1,\vph_2)}.
		\notag
	}
%
%(In the strict sense, the  existence of the resulting states hinges on being normalizable.)
For tube regions, the same relations hold upon replacing $\ta_{1,2}$ with $R_{1,2}$, whereas for rod regions there is no such evolution, since here the boundary consists of one single connected hypersurface.
%Thus there is only an "initial" hypersurface, but no "final" one towards which to evolute the state.
%In other words: evolution of states makes sense only across regions whose boundary has (at least) two disjoint components.

A first consistency condition for the field propagator is the unitarity property,
which assures conservation of the inner product \eqref{eq:inpro_schro} 
under evolution of the states via \eqref{eq:state_propagation_interval},
see \cite{CoOe:unit} for a detailed analysis.
Moreover, it assures that propagating a state from a first hypersurface $\Si_1$ to a second $\Si_2$ via \eqref{eq:state_propagation_interval} and then back again to $\Si_1$, results in the original state.
With the Dirac delta of functional integration, the unitarity property writes as

\bal{	% revised 2016 MAR 26
	\label{eq:unitarity}
	\de(\vph_1,\ki_1)
	& = \int_{K_2}\!\!\!\!\xD{\vph_2}\; 
	\coco{Z\lintval\ta12 (\vph_1,\vph_2)}\;
	Z\lintval\ta12 (\ki_1,\vph_2).
	}
Unitarity \eqref{eq:unitarity} and state propagation \eqref{eq:state_propagation_interval} indicate that complex conjugation yields the inverse of the field propagator, which makes sense when taking into account its  definition \eqref{eq:fieldprop} via the action, plus the fact that reversing the direction of $\ta$ also reverses the sign of the action \eqref{eq:actintreg}.

A second consistency condition for the field propagator is the composition property, which assures that direct propagation from $\Si_{\ta_1}$ to $\Si_{\ta_3}$ and consecutive propagations from $\Si_{\ta_1}$ to $\Si_{\ta_2}$ and then from $\Si_{\ta_2}$ to $\Si_{\ta_3}$ yield the same result:
\bal{
	% revised 2016 MAR 26
	\label{eq:composition}
	Z_{[\ta_1,\ta_3]}(\vph_1,\vph_3)
	= \int_{K_2}\!\!\!\! \xD\vph_2\;
	Z_{[\ta_1,\ta_2]}(\vph_1,\vph_2)\,  
	Z_{[\ta_2,\ta_3]}(\vph_2,\vph_3).
	}
For the field propagators of two tube regions, the same relation holds replacing $\ta_{1,2}$ with $R_{1,2}$.
The composition property \eqref{eq:composition} of the field propagators also assures the gluing property of the amplitudes (T5b) in \cite{Oe:hol}, when gluing together two interval regions or two tube regions, wherein the gluing anomaly factor has value one in both cases.
When gluing together a rod region $M_{R_1}$ with a tube region $M_{[R_1,R_2]}$, then the gluing property (T5b) is assured instead by the following composition property (the gluing anomaly has value one here, too):
\bal{
	% revised 2016 MAR 27
	\label{eq:compositionhyp}
	Z_{R_2}(\vph_2)
	= \int_{K_1}\!\!\!\! \xD\vph_1 \;
	Z_{R_1}(\vph_1)\,
	Z_{[R_1,R_2]}(\vph_1,\vph_2).
	}
\section{Free theory}
\label{sec:freetheory}
\noindent
First, we consider the quantum theory of a free scalar field in interval, tube and rod regions. 
All quantities related to the free theory carry the label 0.
We start with the expression of the free field propagators associated to the different regions. 
Then we define the vacuum and coherent states, and finally compute the free amplitude for the coherent states in the regions of interest.
Since the tube region can be seen as a special case of an interval region (just use spherical coordinates, and partially discrete momenta), we shall focus on interval and rod regions.
%
% ===================================================================
%
\subsection{Field propagators}
\label{sec:freeprops} 
\noindent
The free field propagator \eqref{eq:propinterval} for the interval region $M_{[\ta_1,\ta_2]}$ can be evaluated by shifting the integration variable by a classical solution $\phi\ltx{cl}$ which matches the boundary configurations $\vph_1$ and $\vph_2$ at $\ta =\ta_1$ and $\ta=\ta_2$ respectively.
As in Section~\ref{sec:_classical}, we use a superscript 0 for the free theory.
As usual in path integration, the measure is assumed to be translation-invariant, resulting in
\bal{
	% clear
	Z^0_{[\ta_1,\ta_2]}(\vph_1, \vph_2) 
	= \int_{\begin{matrix}
			\scriptstyle	\phi|_{\Sigma_1}=\vph_1 \\ 
			\scriptstyle \phi|_{\Sigma_2}=\vph_2
			\end{matrix}
			} 
	\xD \phi \; \eu^{\im S^0_{[\ta_1,\ta_2]}(\phi)} 
	= \mc N^{Z,0}_{[\ta_1,\ta_2],} \, 
	\eu^{\im S^0_{[\ta_1,\ta_2]}(\vph_1, \vph_2)}, 
}
wherein the free action $S^0_{[\ta_1,\ta_2]}(\vph_1, \vph_2)$ is additive and given by \eqref{eq:actev} respectively \eqref{eq:actintreg2}, while the normalization factor is formally given by
\bal{
	% clear
	\Norm[Z,0]{[\ta_1,\ta_2]} 
	= \int_{\phi|_{\Sigma_1}=\phi|_{\Sigma_2}=0} \xD \phi \; 
	\eu^{\im S^0_{[\ta_1,\ta_2]}(\phi)}.
}
%
%In order not to run out of letters we shall denote all normalization factors by $\mc N$ with some distinguishing labels attached.
Applying the same technique, the free field propagator in the rod region $M_R$ results to be
\bal{
	% clear
	\label{eq:_free_fieldprop_R}
	Z^0_{R}(\vph_R) 
	= \Norm[Z,0]{R} \, \eu^{\im S^0_{R}(\vph_R)},
}
with $S^0_{R}(\vph_R)$ from \eqref{freeaction_rod_boundary}.
Explicit expressions for the normalization factors $\Norm[Z,0]{[\ta_1,\ta_2]}$ and $\Norm[Z,0]{R}$ are obtained below.
%
% ================================================================
%
\subsection{Vacuum state}
\label{sec:vacuum} 
\noindent
According to the axioms of the GBF, a vacuum state $\ps^0_{\Si_\ta}\in\cH_{\Si_\ta}$ is associated to each hypersurface $\Si_\ta$ (the label 0 here indicates the vacuum, not the free theory).
As before, $\ta$ denotes the foliation parameter, $\Si_\ta$ a leaf of the foliation (backwards oriented), and $\vc x$ coordinates on $\Si_\ta$.
We assume that the wave functional describing the vacuum state on $\Si_\ta$ has the form of a Gaussian:
\bal{	% clear
	\label{zzz_SKG_SF_vacuum_09}
	\psSar[0]{\Si_\ta}\vph = \Norm[\text{S},0]{\Si_\ta}\;
	\exp\, \biiglrr{\!-\frac12  \int_{\Si_\ta}\!\! \xd^3 x\;
		\vph(\vc x)\,  \biglrr{A_{\Si_\ta}\vph}\! (\vc x)
		}	\;,
	}
wherein $A_{\Si_\ta}$ is called the vacuum operator.
The superscript S stands for Schr\"odinger picture.
$\smash{\Norm[\text{S},0]{\Si_\ta}}$ is another real 
normalization factor satisfying the condition
\bal{% revised 2016 MAR 28
	\label{eq:normvac}
	\abss{\Norm[\text{S},0]{\Si_\ta}}^{-2} 
	= \det \biiglrr{ \tfracw{A_{\Si_\ta} + \coco{A_{\Si_\ta}}}
										{2 \pi} 
							}^{\!-1/2}
	:= \int_{K_\Si}\!\!\! \xD\vph\;
		\exp\biiglrr{-\frac12 \int_{\Si} \xd^3x\; \vph(\vc x)\,
						\biglrr{ (A_{\Si_\ta}\!\!+\!\coco{A_{\Si_\ta}})\, \vph
							}(\vc x)
						},
	}
which assures $\inpro{\psS[0]{\Si_\ta}}{\psS[0]{\Si_\ta}} = 1$.
Applying \eqref{eq:state_propagation_interval} and \eqref{eq:inpro_schro}, 
we see that this is already enough to make the free interval 
amplitude~\eqref{eq:amplitude_general_interval} of the vacuum state 
have value one, thereby fulfilling Vacuum Axiom (V5) in \cite{Oe:GBQFT, Oe:hol}.
This condition later helps in fixing $\Norm[Z,0]{[\ta_1,\ta_2]}$ 
and $\Norm[Z,0]{R}$. Since $\ps^0_{\Sibar_\ta} = \coco{\ps^0_{\Si_\ta}}$, 
we have $A_{\Sibar_\ta} = \coco{A_{\Si_\ta}}$.
The general form of the vacuum operator $A_{\Si_\ta}$ has been derived in \cite{Co:vac} from the condition of vacuum conservation under free propagation \eqref{eq:state_propagation_interval} with \eqref{eq:propinterval} and \eqref{eq:actev}, that is:
\bal{
	% clear
	\label{eq:vacuum_conservation}
	\ps^{\text{S},0}_{\Si\ta_2}(\vph_2)
	& \overset!= \int_{K_1}\!\!\!\! \xD\vph_1\;
	\ps^{\text{S},0}_{\Si\ta_1}(\vph_1)\,
	Z^0_{[\ta_1,\ta_2]}(\vph_1,\vph_2),
	}
wherein both $\ps^{\text{S},0}_{\Si\ta_{1,2}}$ have the form \eqref{zzz_SKG_SF_vacuum_09}.
% We use the symmetry of $A$ in several calculations,
% the most important being the derivation of the operator's form
% \eqref{eq:vacuum} from \eqref{eq:vacuum_conservation}.
Moreover, the vacuum operator is assumed to be symmetric with respect 
to the inner product $\inpro{\vph}{\ti\vph}$, like the $X\htx{a}(\ta)$ 
and $X\htx{b}(\ta)$, see the end of Section \ref{sec:interval}.
In our notation the result of these conditions writes as
\bal{	% revised 2016 MAR 31
	\label{eq:vacuum}
	A_{\Si_\ta} = \!-\!\im \si\, 
	\sqrt{\abs{g^{(3)}g^{\ta\ta}}_\ta}\,
	\fracw{\coco{(\del_\ta\Up)(\ta)}}{\coco{\,\Up(\ta)\,}} \,,
	}
wherein $g^{(3)}_{\Si_\ta}$ is again the three-metric induced on $\Si_\ta$, and we have introduced the operator
\bal{	% clear
	\label{eq:upsilon}
	\Up(\ta) 
	& \defeq c\htx a X\htx a(\ta) + c\htx b X\htx b(\ta),
}
which acts on the modes $U\lvc k(\vc x)$ as
\bal{	% clear
	\label{eq:upsilon_k}
	\Up(\ta) U\lvc k (\vc x) 
	& = \Up\lvc k (\ta)\, U\lvc k (\vc x) = \left( c\htx a\lvc k X\htx a\lvc k (\ta) 
		 + c\htx b \lvc k X\htx b\lvc k(\ta) \right) U\lvc k (\vc x).
	}
Therein, $c\htx a$ and $c\htx b$ are linear operators defined by their complex eigenvalues $\smash{c\htx a\lvc k}$ and $\smash{c\htx b\lvc k}$ when acting on the $\smash{U\lvc k (\vc x)}$.
The choice of these eigenvalues determines the vacuum operator and hence the vacuum state.
However, in order for the vacuum operator $A_{\Si_\ta}$ to be symmetric, we need the reflection properties $c\htx a\lvc k = c\htx a_{\!-\! \vc k}$ and $c\htx b\lvc k = c\htx b_{\!-\! \vc k}$, which together with \eqref{eq:Xab_k_reflect} induce the reflection property $\Up_{-\vc k}(\ta) = \Up_{\vc k}(\ta)$.
(For an equal time-hyperplane in Minkowski spacetime we have $\Up_{\vc k}(t) = 2E_{\vc k}\,(2\pi)^3\, \eu^{-\im E_{\vc k} t}$, with $A_{\Si_t} = \sqrt{-\vc\del^2+m^2}$\,).
A relation that is often useful in calculations is $\biglrr{\coco{\Up}\del_\ta\!\Up - \Up\,\coco{\del_\ta\!\Up}}(\ta) = 2\im\,\Impart(\coco{c\htx a}c\htx b)\,\mc W(\ta)$.
In order to prevent the exponential in the vacuum state from diverging, the real part of the vacuum operator 
\bal{	% revised 2016 MAR 28
	\label{zzz_SKG_SF_vacuum_101_5}
	\AR_{\Si_\ta} 
	& \defeq \tfrac12 \bglrr{A_{\Si_\ta} \!+ \coco{A_{\Si_\ta}\!}\,}
	= -\si \,\sqrt{\abs{g^{(3)}g^{\ta\ta}}_\ta}\,
		\fracw{\Impart (\coco{c\htx a} c\htx b)\,\mc W (\ta)}{\abs{\Up(\ta)}^2}
	}
must be positive. This implies the positivity condition
\bal{	% clear
	\label{zzz_SKG_SF_vacuum_102}
	-\si\,\Impart\biglrr{\coco{c\htx a_{\vc k}}c\htx b_{\vc k}}\,
	\mc W\lvc k (\ta) 
	> 0,
	}
which in particular requires that $2\im\, \Impart (\coco{c\htx a_{\vc k}} c\htx b_{\vc k}) = \coco{c\htx a_{\vc k}} c\htx b_{\vc k}- c\htx a_{\vc k} \coco{c\htx b_{\vc k}} \neq 0$.
This implies that $c\htx a_{\vc k}\neq 0\neq c\htx b_{\vc k}$. 
Since $X\htx{a}_{\vc k}(\ta)$ and $X\htx{b}_{\vc k}(\ta)$ never vanish 
at the same $\ta$, we get $\Up_{\vc k}(\ta) \neq 0$.
This makes $\Up(\ta)$ invertible, and hence $A_{\Si_\ta}$ in \eqref{eq:vacuum} and $\AR_{\Si_\ta}$ in \eqref{zzz_SKG_SF_vacuum_101_5} are well defined.
Further, since $\del_\ta X\htx{a}_{\vc k}(\ta)$ and 
$\del_\ta X\htx{b}_{\vc k}(\ta)$ neither vanish at the same $\ta$, 
we also get $\del_\ta \Up_{\vc k}(\ta) \neq 0$.
This makes $A_{\Si_\ta}$ invertible, and $\AR_{\Si_\ta}$ is invertible due to \eqref{zzz_SKG_SF_vacuum_102}.
If \eqref{zzz_SKG_SF_vacuum_102} is fulfilled, then we can also write
\bal{% clear
	\label{eq:AR_absval}
	\AR_{\Si_\ta} & = 
	\sqrt{\abs{g^{(3)}g^{\ta\ta}}_\ta}\,
	\fracw{\abss{\Impart (\coco{c\htx a} c\htx b)\,\mc W (\ta)}}{\abs{\Up(\ta)}^2}\;.
	}
The composition property \eqref{eq:composition} and unitarity property \eqref{eq:unitarity} of the field propagator imply several relations for the various normalization factors, which together fix these normalization factors up to a complex phase.
We here choose different phases than those in \cite{CoOe:unit}, in  particular we choose the vacuum normalization factor to be real as called for in Section IV.B of \cite{Oe:KGtl}. 
A complex normalization factor might obstruct the vacuum state from becoming a real-valued functional, which is required by the GBF's Axioms whenever there is an isometry (connected to the identity) that reverses the orientation of the hypersurface $\Si_\ta$.
(For an example [spatial rotation], see Section IV.B of \cite{Oe:KGtl}, where $\ta=x^1$ is a spatial cartesian coordinate.)
Since this is not the case in generic spacetimes, we shall in general not require the vacuum state to be real-valued, and only avoid possible obstructions to this.
Our normalization factors write,
\bal{	% clear from {zzz_SKG_SF_vacuum_101_5}, {eq:AR_absval}
	\label{zzz_SKG_SF_vacuum_normalization factors_102}
	\Norm[\text{S},0]{\Si_\ta} & =
	\deth[1/4]\biiglrr{\tfrac{2\AR_{\Si_\ta}}{2\pi}}
	= \deth[1/4] 
		\biiglrr{\sqrt{\abs{g^{(3)}g^{\ta\ta}}_\ta}
					\tfracw{\abs{2\,\Impart (\coco{c\htx a} c\htx b)\,\mc W (\ta)}}
								{2\pi\,\abs{\Up(\ta)}^2}
					},
		\\
	% revised 2016 MAR 30
	\label{zzz_SKG_SF_vacuum_normalization factors_100}
	\Norm[Z,0]{\intval\ta12}
	& = \deth[1/2] \biiglrr{
		\tfracw{\im W^{12}_{[\ta_1,\ta_2]}}{2\pi}
		\tfracw{\abs{\Up(\ta_1)}}{\coco{\,\Up(\ta_1)\,}}
		\tfracw{\coco{\,\Up(\ta_2)}\,}{\abs{\Up(\ta_2)}}
		},
		\\
	% revised 2016 MAR 30
	\label{zzz_SKG_SF_vacuum_normalization factors_110}
	\Norm[Z,0]{R} & = \deth[1/4] \biiglrr{-
		\sqrt{\abs{g^{(3)}g^{rr}}_R}
		\tfracw{(c\htx b)^2}
				{2\pi\,\abs{2\Impart (\coco{c\htx a} c\htx b)}}
		\tfracw{\abs{\mc W(R)}}{(X\htx a(R))^2}
		\tfracw{\coco{\,\Up(R)\,}}{\Up(R)}
		}.
	}
%
% ==================================================================
%
\subsection{Coherent states}
\label{sec:coherent}
\noindent
In this section we introduce coherent states since they have been useful in \cite{CoOe:smatrix,Co:dS} to compute amplitudes.
In the Schr\"odinger representation a coherent state living on $\Si_\ta$ is determined by a \emph{complex} function $\et(\vc x)$ on $\Si_\ta$. 
We can view it as a complexified configuration and call it the characteristic function of the coherent state. 
As always, $\Si_\ta$ is oriented backwards.
%
% ====================================================================
%
\subsubsection{Schr{\"o}dinger picture} 
\noindent
Here the coherent states depend on $\ta$, and are evoluted from a leaf $\Si_\ta$ to another by the free field propagator.
In the next subsection we advance to the Dirac (interaction) picture,
in which the coherent states are invariant under free evolution.
We define a coherent state to map a configuration $\vph$ on $\Si_\ta$ to a complex number, as given by a slightly modified version of (24) in \cite{CoOe:smatrix}:
\bal{	% clear
	\label{zzz_SKG_SF_coherent_100}
	\psS[\et]{\Si_\ta}(\vph) 
	& = \Norm[\text{S},\et]{\Si_\ta}\;
	\exp \,\biiglrr{\int_{\Si_\ta}\!\!\! \xd^3 x
						\sqrt{\abs{g^{(3)}g^{\ta\ta}}_\ta}\,
						\vph(\vc x)\, \et(\vc x)
						}\;
	\psS[0]{\Si_\ta}(\vph)
		\\
	% momentum space expression, included only 
	% for easier comparison with Minkowski papers
	% revised 2016 APR 01
	\label{zzz_SKG_SF_coherent_101}
	& = \Norm[\text{S},\et]{\Si_\ta}\;
	\exp \,\biiglrr{\int\!\xd^3 k\; 
						\vph_{\vc k}\, \ti w_{\vc k}(\ta)\, \et_{-\vc k}	
						}\;
	\psS[0]{\Si_\ta}(\vph).
	}
The second line can be obtained from the first using the 
expansion \eqref{eq:modeinterval} on $\Si_\ta$.
However, since $\et(\vc x)$ is complex, in general $\et_{- \vc k}\neq\coco{\et_{\vc k}}$.
As suggested by the notation, the vacuum state $\psS[0]{\Si_\ta}$ is precisely the coherent state characterized by the identically vanishing function $\et(\vc x) \equiv 0$.
Requiring a coherent state to be normalized: $\inproo{\psS[\et]{\Si_\ta}}{\psS[\et]{\Si_\ta}}= 1$, lets us choose the normalization factor to be%
%
%\footnote{\eqref{zzz_SKG_SF_coherent_110} is merely 
%	one possible choice, fixed up to a complex phase factor 
%	$\exp (\im \al_\et)$ with $\al_\et \in\reals$.
%	Our choice is convenient, because it relates nicely 
%	to coherent states in the holomorphic representation.
%	For the vacuum, a similar phase factor is fixed to 1,
%	because it makes sense to require the vacuum state to be real, 
%	see Section IV.B in \cite{Oe:KGtl}.
%	However, being necessarily real is not justified for other states.
%	\eqref{eq:K1K2} does not fix the phase factor, 
%	because \eqref{eq:eta1eta2} relates $\et_1$ and $\et_2$.
%	We see no physical criterion to fix this ambiguity, and
%	since the amplitude map is linear, the amplitudes inherit
%	these phase factors.
%	This shows that they cause no problem, because GBF probabilities
%	depend only on the absolute value squared of the amplitudes
%	\cite{Oe:GBQFT,Oe:probgbf}, similar to the standard probabilities
%	obtained from Born's rule.
%	}
%

\bal{	 % revised 2016 APR 01
	\label{zzz_SKG_SF_coherent_110}
	\Norm[\text{S},\et]{\Si_\ta} 
	& = \exp\,\biiglrr{-\frac12\int_{\Si_\ta}\!\!\!\xd^3 x
									\sqrt{\abs{g^{(3)}g^{\ta\ta}}_\ta}\,
		 							\et(\vc x)\; \KS_{\Si_\ta} 
									(\et\!+\!\coco\et)(\vc x)
									},
		\\
	% revised 2016 APR 01
	\KS_{\Si_\ta} 
	& \defeq \tfrac12\,\sqrt{\abs{g^{(3)}g^{\ta\ta}}_\ta}\,
		/\AR_{\Si_\ta},
	}
wherein $\KS_{\Si_\ta}$ is symmetric and real.
The inner product of two coherent states can be calculated using a shift of integration variable $\vph\to\vph + \KS_{\Si_\ta}(\coco\et\!+\!\ki)$ and yields
\bal{		% revised 2016 APR 01
	\label{zzz_SKG_SF_coherent_150}
	\inproo{\psS[\et]{\Si_\ta}}{\psS[\ki]{\Si_\ta}}
	= \exp\, \biggl(\, \int_{\Si_\ta}\!\!\!\xd^3 x\;
						\sqrt{\abs{g^{(3)}g^{\ta\ta}}_\ta}\,
					\biiglrr{\coco\et\,\KS_{\Si_\ta} \ki
						-\tfrac12 \coco\et\, \KS_{\Si_\ta} \et
						-\tfrac12 \coco\ki \,\KS_{\Si_\ta} \ki
						}(\vc x)\!
			\biggr).
	}
The coherent states fulfill the completeness relation
\bals{	% clear
	\One 
	& = \Norm[\text{S},\One]{\Si_\ta} \int\!
		\xD{\vph}\, \xD{\coco\vph}\;
	\kett{\psS[\vph]{\Si_\ta}}\,\braa{\psS[\vph]{\Si_\ta}},
		\\
	% revised 2016 APR 01
	\biglrr{\Norm[\text{S},\One]{\Si_\ta}}^{-1} 
	& = \int\!	\xD{\vph}\, \xD{\coco\vph}\;
	\exp\, \biiglrr{\!-\!\! \int_{\Si_\ta}\!\!\!\xd^3 x
					\sqrt{\abs{g^{(3)}g^{\ta\ta}}_\ta}\,
				 	\coco{\vph(\vc x)}\,	\KS_{\Si_\ta} \vph(\vc x) 
					}.
	}
This normalisation constant is real and can be computed from the following completeness relation using the shifts $\vph\to \vph\!+\!\ki$ and $\coco\vph \to \coco\vph+\coco\et$:
\bal{	% revised 2016 APR 01
	\label{zzz_SKG_SF_coherent_180}
	\inproo{\psS[\et]{\Si_\ta}}{\psS[\ki]{\Si_\ta}}
	= \inproo{\psS[\et]{\Si_\ta}}{\One\,\psS[\ki]{\Si_\ta}}
	= \Norm[\text{S},\One]{\Si_\ta} \int_{K^{\complex}_{\ta}}\!\!\!
	\xD{\vph}\, \xD{\coco\vph}\;
	\inproo{\psS[\et]{\Si_\ta}}{\psS[\vph]{\Si_\ta}}\;
	\inproo{\psS[\vph]{\Si_\ta}}{\psS[\ki]{\Si_\ta}}\;.
	}
Coherent states remain coherent under free evolution  \eqref{eq:state_propagation_interval}.
That is, propagating a coherent state with characteristic function $\et_1(\vc x)$ results in a new coherent state with a generically different $\et_2(\vc x)$ as in
\bal{
	% clear
	\label{eq:state_prop_coherent_interval}
	\psS[\et_2]{\Si\ta_2}(\vph_2)
	= \int\!\! \xD\vph_1\, \psS[\et_1]{\Si\ta_1}(\vph_1)\;
	Z^0_{[\ta_1,\ta_2]}(\vph_1,\vph_2).
}
We can evaluate this by shifting $\vph_1 \to \vph_1+[A_{\Si_{\ta_1}}\!\!-\!\im W^{11}\lintval\ta12]^{-1} (w\ti w\et_1 \!+\! \im W^{12}\lintval\ta12\vph_2)$, leading to the following two relations which determine 
(the characteristic function and the normalization factor of) 
the coherent state $\psS[\et_2]{\Si\ta_2}$:

\bal{	% revised 2016 APR 01
	\label{eq:eta1eta2}
	\et_2(\vc x) 
	& = \tfrac{\ti w(\ta_1)}{\ti w(\ta_2)}
		\tfracw{\im W_{[\ta_1,\ta_2]}^{12}}
					{A_{\ta_1} \!- \im W_{[\ta_1,\ta_2]}^{11}}
		\et_1 (\vc x)\,  
	= \tfracw{\ti w(\ta_1)\,\coco{\Up(\ta_1)}}
				{\ti w(\ta_2)\,\coco{\Up(\ta_2)}}
		\et_1 (\vc x),
		\\
	% revised 2016 APR 03
	\label{eq:K1K2}
	\Norm[\text{S},\et_2]{\Si\ta_2}
	& = \Norm[\text{S},\et_1]{\Si\ta_1}\,
	\exp \biiglrr{\frac12\int_{\Si_\ta}\!\!\!\xd^3 x\;
					\et_1(\vc x) \tfracw{w^2(\vc x)\,\ti w^2(\ta_1)}
						{A_{\ta_1}\!\!-\!\im W^{11}\lintval\ta12}
					\et_1(\vc x) 
					}.
	}
Substituting \eqref{eq:eta1eta2} in \eqref{eq:K1K2} and using expressions \eqref{eq:vacuum} and \eqref{eq:W2}, one can verify with simple algebra that relation \eqref{eq:K1K2} is indeed satisfied.
\eqref{eq:eta1eta2} tells us, that the characteristic functions $\et_1$ of the original state and $\et_2$ of the evoluted state are related through

\bal{		% revised 2016 APR 01
	\label{zzz_SKG_SF_coherent_240}
	\ti w(\ta_2)\,\coco{\Up(\ta_2)}\,\et_2(\vc x)  
	& = \ti w(\ta_1)\,\coco{\Up(\ta_1)}\, \et_1(\vc x),
		&
	% clear
	\ti w_{\vc k}(\ta_2)\,\coco{\Up\lvc k(\ta_2)}\,\et_{2,\vc k}  
	& = \ti w_{\vc k}(\ta_1)\,\coco{\Up\lvc k(\ta_1)}\, \et_{1,\vc k}.
	}
This confirms that the vacuum state is preserved under free evolution, since it has the characteristic function $\et(\vc x) \equiv 0$.
%
% ===========================================================
%
\subsubsection{Dirac picture (interaction picture)}
\noindent
With \eqref{zzz_SKG_SF_coherent_240} in mind, 
we define our Dirac picture coherent states as
\bal{% clear
	\psD[\et]{\Si_\ta} 
	& \defeq \psS[(\ti w(\ta)\coco{\Up(\ta)})^{-1}\et]{\Si_\ta}\;,
	\\
	% clear
	\label{zzz_SKG_SF_coherent_300}
	\psD[\et]{\Si_\ta}(\vph) & = \Norm[\text{D},\et]{\Si_\ta}\;
	\exp \,\biiglrr{\int\!\xd^3 x\; \vph(\vc x)
						\tfracw{w(\vc x)}{\coco{\Up(\ta)}} \et(\vc x)
						}\;
	\psS[0]{\Si_\ta}(\vph).
	}
We remark that the Dirac picture coherent states \emph{are not} $\ta$-independent, but rather related through evolution with the free field propagator: Evoluting $\psD[\et]{\Si\ta_1}$ as in \eqref{eq:state_prop_coherent_interval}, we obtain $\psD[\et]{\Si\ta_2}$. 
That is, the evoluted state now has the \emph{same} characteristic function $\et$ as the initial state.
Again, the vacuum $\psD[0]{\Si_\ta}$ is the coherent state characterized by $\et(\vc x) \equiv 0$ and thus $\psD[0]{\Si_\ta}(\varphi)= \psS[0]{\Si_\ta}(\varphi)$.
All relations of the Schr\"odinger picture are easily adapted to the Dirac picture.
The normalization \eqref{zzz_SKG_SF_coherent_110} now writes as
\bal{		% revised 2016 APR 02
	\label{zzz_SKG_SF_coherent_310}
	\Norm[\text{D},\et]{\Si_\ta} 
	= \exp\, \biiiglrc{\!-\!\tfrac12\!\int_{\Si_\ta}\!\!\!\xd^3 x\;
			\biiglrr{\et(\vc x) 
				\tfracw{w(\vc x)\KS_{\Si_\ta}/\ti w(\ta)}{\coco{\Up(\ta)}^2}
				\et(\vc x)  +  \et(\vc x)
				\tfracw{w(\vc x)\KS_{\Si_\ta}/\ti w(\ta)}{\abs{\Up(\ta)}^2} 
				\coco{\et(\vc x)}
				}\!
		},
	}
ensuring $\inproo{\psD[\et]{\Si_\ta}}{\psD[\et]{\Si_\ta}}= 1$.
The inner product \eqref{zzz_SKG_SF_coherent_150} becomes
\bal{		% clear
	\label{zzz_SKG_SF_coherent_350}
	\inproo{\psD[\et]{\Si_\ta}}{\psD[\ki]{\Si_\ta}}
	& = \exp\, \biggl(\, \int_{\Si_\ta}\!\!\!\xd^3 x\; 
		\biiglrr{ \coco\et\,w\KD \ki
			-\tfrac12 \coco\et\,w\KD \et
			- \tfrac12 \coco\ki \,w\KD \ki
			}(\vc x)\!
		\biggr).
	}
with the $\ta$-independent (see below \eqref{SKG_structures_KleiGo_33}), real, symmetric operator $\KD \defeq \KS_{\Si_\ta} \abs{\Up(\ta)}^{-2}\!\! / \ti w(\ta)$ given by:
\bal{	% revised 2016 APR 03
	\KD  
	= & \abss{2\Impart (\coco{c\htx a} c\htx b)\, 
					\ti w(\ta)\,\mc W(\ta)}^{-1},
		&
	% clear
	\KD \lvc k 
	= & \abss{2\Impart\biglrr{\coco{c\htx a_{\vc k}}c\htx b_{\vc k}}\, \ti w_{\vc k}(\ta)\, \mc W\lvc k(\ta)}^{-1}.
	}
The coherent Dirac states fulfill the completeness relation
\bals{	% clear
	\One & = \Norm[\text{D}]\One 
	\int\! \xD{\vph}\, \xD{\coco\vph}\;
	\kett{\psD[\vph]{\Si_\ta}}\,\braa{\psD[\vph]{\Si_\ta}},
		\\
	% clear
	\biglrr{\Norm[\text{D}]\One}^{-1} 
	& = \int\! \xD{\vph}\, \xD{\coco\vph}\;
		\exp \biiglrr{ -\!\!\int_{\Si_\ta}\!\!\!\xd^3 x\;
		\coco{\vph(\vc x)}\, w(\vc x)\KD \, \vph(\vc x)}.
	}
This real normalization factor can be computed
from the completeness relation
\bal{		% clear
	\label{zzz_SKG_SF_coherent_380}
	\inproo{\psD[\et]{\Si_\ta}}{\psD[\ki]{\Si_\ta}}
	= \inproo{\psD[\et]{\Si_\ta}}{\One\,\psD[\ki]{\Si_\ta}}
	= \Norm[\text{D}]\One \int\!\xD{\vph}\,\xD{\coco\vph}\;
	\inproo{\psD[\et]{\Si_\ta}}{\psD[\vph]{\Si_\ta}}\;
	\inproo{\psD[\vph]{\Si_\ta}}{\psD[\ki]{\Si_\ta}}.
	}
Analogous formulas with $\ta$ replaced by $r$ hold for coherent states 
defined on hypercylinders $\Si_r$.
%
%==================================================================
\subsection{Free amplitudes}
\label{sec:freeampl}
\noindent
After these preparations, we can now explicitly compute the free amplitude for coherent states in the Dirac picture. 
We denote the amplitude map resulting from Schr\"odinger-Feynman Quantization (SFQ) always by $\roS{}$, both for states $\psS{}$ in the Schr\"odinger picture and $\psD{}$ in the Dirac picture.
In SFQ, the amplitude map associated to a region is always given by \eqref{eq:amplitude_general}, of which the interval and tube amplitudes \eqref{eq:amplitude_general_interval} and rod amplitudes \eqref{eq:amplitude_general_rod} are special cases.

For an interval region $M\lintval\ta12$, we now calculate the free amplitude for a boundary state $\psD[\et]{\Si\ta_1}\otimes \coco{\psD[\ki]{\Si\ta_2}}$.
The first coherent state is defined by the complex function $\et$ at $\ta_1$ and the second by $\ki$ at $\ta_2$.
Here the boundary is oriented outwards of the region, that is: $\Si_{\ta_1}$ canonically backwards and $\Sibar_{\ta_2}$ forwards (causing the complex conjugation of the state on $\Sibar_{\ta_2}$).
According to \eqref{eq:amplitude_general_interval} the interval's free amplitude results to be

\bal{		% clear
	\roS[0]{\intval\ta12} 
		\biiglrr{\psD[\et]{\Si\ta_1}\!\otimes\! \coco{\psD[\ki]{\Si\ta_2}}\,}
	& = \int_{K_1}\!\!\!\!\xD\vph_1\!
		\int_{K_2}\!\!\!\!\xD\vph_2\;
		\psD[\et]{\Si\ta_1}(\vph_1) \;
		\coco{\psD[\ki]{\Si\ta_2}(\vph_2)}\;
		Z^0\lintval\ta12(\vph_1,\vph_2)
%		\notag
%		\\
%	%	clear
%	& = \int_{K_2}\!\!\!\!\xD\vph_2\;
%		\psD[\et]{\Si\ta_2}(\vph_2) 
%		\;\coco{\psD[\ki]{\Si\ta_2}(\vph_2)}\;
%		%
		= \inprooo{\psD[\ki]{\Si\ta_2}}
											{\psD[\et]{\Si\ta_2}}
		\notag
		\\
	% clear
	\label{zzz_SKG_SF_free_amplitude_11}
	& = \exp\, \biggl(\, \int\!\! \xd^3 x\;
		\biiglrr{ \et\,w\KD \coco\ki
			-\tfrac12 \coco\et\,w\KD \et 
			- \tfrac12 \coco\ki \,w\KD \ki
			}(\vc x)\!
		\biggr).
	}
%
%This relates to bra-ket notation via $\roS[0]{\intval\ta12} 
%\biglrr{\psD[\et]{\Si\ta_1}\!\otimes\! \coco{\psD[\ki]{\Si\ta_2}}\,} =
%\braopkett{\psD[\ki]{\Si\ta_1}}{\mc S^0}{\psD[\et]{\Si\ta_2}}$.
Since the operator $\KD$ is independent of $\ta$, so is the whole free amplitude (this is to be expected, since we are considering the free evolution of states in the interaction picture).
Therefore it is trivial to take the limit $\ta_1\to\!-\!\infty$ and $\ta_2\to\!+\!\infty$, and we interpret this limit of the amplitude as the S-matrix.
IF $\ta$ is a time function, then this is the usual S-matrix.
Since $\KD$ essentially determines the free amplitude, we call it the free amplitude operator.
For $\et = \ki$, the integrand in \eqref{zzz_SKG_SF_free_amplitude_11} vanishes.
Thus the amplitude of an initial state $\psD[\et]{\Si\ta_1}$ and a final state $\psD[\et]{\Si\ta_2}$, which is the freely evoluted initial state, has value one as expected for a Dirac state in free theory.
For tube regions $M\lintval R12$, from their analogue of \eqref{eq:amplitude_general_interval} we obtain the analogue expression
(the boundary is again oriented outwards of the region, that is: $\smash{\Si_{R_1}}$ inwards and $\smash{\Si_{R_2}}$ outwards with respect to $r$):
\bal{		% clear
	\label{zzz_SKG_SF_free_amplitude_11_tube}
	\roS[0]{\intval R12} \biiglrr{\psD[\et]{\Si_{R1}}
		\!\otimes\! \coco{\psD[\ki]{\Si_{R2}}}\,}
	& = \exp\, \biggl(\, \int\!\! \xd t\, \xd\Om\;
		\biiglrr{ \et\,w\KD \coco\ki
			-\tfrac12 \coco\et\,w\KD \et 
			- \tfrac12 \coco\ki \,w\KD \ki
			}(t,\Om)\!
		\biggr).
	}
We also calculate the free amplitude for a boundary state of a rod region $M_R$. 
The boundary is oriented outwards of the region, which is the same as outwards with respect to $r$, and accounts for the complex conjugation. 
The integral over field configurations in the expression of the rod amplitude \eqref{eq:amplitude_general_rod} for a coherent state \eqref{zzz_SKG_SF_coherent_300} can be evaluated by shifting the integration variable as $\vph \to \vph + w_{\om l m_l\!}(t,\Om)\,\Up(R)^{-1} (\coco{A_R}-\im W_R)^{-1}\, \coco{\ki}$, resulting in the expression

\bal{		% revised 2016 APR 03
	\label{zzz_SKG_free_amplitude_100}
	\roS[0]{\Si_R}\biiglrr{\,\coco{\psD[\ki]{\Si_R} }\, } 
	& = \exp\, \biiglrr{-\frac12 \int_{\Si_R}\!\!\!\!
			\xd t\,  \xd\Om\; \biglrr{
					\coco\ki \tfracw{\coco{c\htx b}}{c\htx b}w\KD \coco\ki
					+ \coco{\ki}\, w\KD \ki
					}(t,\Om)
			}.
	}
Since $\KD$ and $c\htx b$ are independent of $R$, so is the whole free amplitude, and the limit $R \to \!+\!\infty$ is trivial.
We interpret this limit as the radial S-matrix.
%
% =================================================================
%
\section{Interacting theory}
\label{sec:inttheory}
\noindent
Following \cite{CoOe:smatrix,Co:dS}, as an intermediate step toward the general interacting theory, we consider now the interaction of the scalar field with a real source field $\mu$. 
The corresponding action carries the label $\mu$,
\bal{
	% clear
	\label{eq:actionsrc}
	S^\mu_M(\phi) = S^0_M(\phi) 
		+ \int_M\!\! \xd ^4 x \, \sqrt{|g(x)|} \; \mu(x) \, \phi(x).
}
We assume that the source $\mu$ is confined in the interior of the region $M$, that is: $\mu(x) =0$ for $x \in \del M$ and $x \notin M$. 
The corresponding propagator is evaluated with the technique applied in Section \ref{sec:freeprops} (shifting the integration variable of the path integral by a classical solution of the free theory, which matches the field configurations on the boundary $\del M$). 
The amplitude in the presence of the source field $\mu$ in interval and rod regions is calculated in the following two subsections. 
The amplitudes for general interactions are then obtained in the last subsection using functional derivative techniques.
%
% =================================================================
%
\subsection{Source field on interval region}
\label{sec:intamplinterval}
\noindent
For interval regions, the field propagator with source field can be expressed in terms of the free one as
\bal{
	% revised 2016 APR 05
	\label{eq:propmu}
	Z^\mu_{[\ta_1,\ta_2]}(\vph) 
	= \fracw{\Norm[Z,\mu]{[\ta_1,\ta_2]}}{\Norm[Z,0]{[\ta_1,\ta_2]}}\, 
		Z^0_{[\ta_1, \ta_2]}(\vph_1,\vph_2)\, 
		\exp \biiglrr{ \im \int\!\! \xd^3 x  \, 
							\biglrr{\mu_1 \, \vph_1 + \mu_2 \, \vph_2
										}(\vc x )
			 				},
}
wherein we have introduced the real quantities
\bal{
	% revised 2016 APR 05
	\mu_1(\vc x ) 
	& \defeq \int_{\ta_1}^{\ta_2}\!\! \xd \ta \, 
		\sqrt{|g(\ta, \vc x )|}
		\fracw{\mc W^{00}(\ta,\ta_2)}{\mc W^{00}(\ta_1,\ta_2)}
		\mu(\ta, \vc x ), 
		&
	\mu_2(\vc x ) 
	& \defeq \int_{\ta_1}^{\ta_2} \!\!\xd \ta \, 
		\sqrt{|g(\ta, \vc x )|}
		\fracw{\mc W^{00}(\ta_1,\ta)}{\mc W^{00}(\ta_1,\ta_2)}
		\mu(\ta, \vc x ),
	}
and the normalization factor $\Norm[Z,\mu]{[\ta_1,\ta_2]}$ is formally equal to
\bal{
	% revised 2016 APR 05
	\Norm[Z,\mu]{[\ta_1,\ta_2]} 
	= \int_{\phi|_{\ta_1}=\phi|_{\ta_2}=0}\!\!\!\! \xD\phi\; 
		\eu^{\im S^\mu_{[\ta_1,\ta_2]}(\phi)}.
}
The integral therein can be evaluated via shifting $\ph\to \ph+\al$ by a solution $\al(x)$ of the inhomogeneous Klein-Gordon equation 
\bal{% clear
	\label{eq:KG_al12_eq_mu}
	(\si_{00}\Box+m^2)\, \al(x) = \mu(x)
	}
with vanishing boundary conditions $0 = \al(\ta_1,\vc x) = \al(\ta_2,\vc x)$.
Due to \eqref{eq:KG_al12_eq_mu}, we have $S^\mu\lintval\ta12(\ph+\al) =S^0\lintval\ta12(\ph) + \tfrac12 \int_M \xd^4x\,\sqrt{|g(x)|}\,\al(x) \mu(x)$, and the quotient $\Norm[Z,\mu]{[\ta_1,\ta_2]}/\Norm[Z,0]{[\ta_1,\ta_2]}$ becomes
\bal{
	% revised 2016 APR 05
	\label{eq:normmu}
	\frac{\Norm[Z,\mu]{[\ta_1,\ta_2]}}{\Norm[Z,0]{[\ta_1,\ta_2]}} 
	= \exp \biiglrr{ \frac{\im}{2} \int\! \xd^4 x \, 
						\sqrt{|g(x)|}\, \al(x) \, \mu(x) 
						}.
}
With $\te$ the Heaviside step function, we express $\alpha$ as
the following real solution of \eqref{eq:KG_al12_eq_mu}:
%(we are still free to add to $\al$ any homogeneous Klein-Gordon solution $\ti\al$ vanishing at $\ta_1$ and $\ta_2$):
%

\bal{		% revised 2016 APR 06
	\al(\ta,\vc x) \!
% equivalent symmetric form of $\alpha$, left out for brevity
%	& = -\si \int_{\ta_1}^{\ta_2} \!\!\xd \ta' \, 
%		\sqrt{|g(\ta', \vc x )|}
%		\biiiglrr{\te(\ta'-\ta) 
%				\fracw{\mc W^{00}(\ta_1,\ta)\, \mc W^{00}(\ta',\ta_2)}
%					{\mc W^{00}(\ta_1,\ta_2)\,w_{\vc k}(\vc x)\ti w_{\vc k}(\ta)\mc W(\ta)} 
%				+\te(\ta-\ta') 
%				\frac{\mc W^{00}(\ta_1,\ta') \mc W^{00}(\ta,\ta_2)}
%					{\mc W^{00}(\ta_1,\ta_2)\,w_{\vc k}(\vc x)\ti w_{\vc k}(\ta)\mc W(\ta)}
%					} 
%		\mu(\ta',\vc x)
%		\\
	% revised 2016 APR 06
	\label{eq:_alpha12_asym}
	& = -\si\!\! \int_{\ta_1}^{\ta_2}\!\!\!\!\!\! \xd \ta' 
		\sqrt{|g(\ta', \vc x )|}\,
		\biiglrr{\!\!
				\tfracw{\mc W^{00}(\ta_1,\ta)\, \mc W^{00}(\ta',\ta_2)}
					{\mc W^{00}(\ta_1,\ta_2)\,w(\vc x)\ti w(\ta)\mc W(\ta)} 
%		\\
%	& \hspace{20mm}
				\!\!+\te(\ta\!-\!\ta') 
				\tfrac{\mc W^{00}(\ta,\ta')}
					{\mc W^{00}(\ta_1,\ta_2)\,w(\vc x)\ti w(\ta)\mc W(\ta)}
					} 
		\mu(\ta',\vc x).
%		\notag
	}
%
%Checking that $\al$ vanishes at $\ta_{1,2}$ is rather straightforward.
%For verifying that it is a solution of \eqref{eq:KG_al12_eq_mu}, note that $\Box$ is sensitive to $\ta$, not to $\ta'$.
%Hence we can view $\mu(\ta',\vc x)$ as reduced initial data, and thus $X\htx{a,b}(\ta)\,\mu(\ta',\vc x)$ are homogeneous solutions, and so are linear combinations of them.
%Acting with the Klein-Gordon operator on \eqref{eq:_alpha12_asym}, it is the $\te$-term which generates the $\mu$-source, while the other term is homogeneous (since $\ti w_{\vc k}(\ta)\mc W(\ta)$ is independent of $\ta$).
It is straightforward to show that the field propagator with source \eqref{eq:propmu} continues to satisfy the unitarity property \eqref{eq:unitarity}.
That is, evolution of states \eqref{eq:state_propagation_interval} with $Z^\mu\lintval\ta12$ still conserves the inner product.
The composition properties \eqref{eq:composition} and \eqref{eq:compositionhyp} continue to hold as well, which can be seen  from the definitions \eqref{eq:propinterval} and \eqref{eq:proprod}.
%As expected for a theory with interaction, the vacuum ceases to be conserved under evolution with $Z^\mu\lintval\ta12$, and ceases to have unit amplitude: $\ro^{\text{S},\mu}\lintval\ta12(\psS[0]{\ta_1}\otimes\coco{\psS[0]{\ta_2}}) \neq 1$.
We now apply the field propagator \eqref{eq:propmu} to calculate the amplitude for the boundary state $\psD[\et]{\ta_1} \otimes \coco{\psD[\ki]{\ta_2}}$ in the presence of the source $\mu$,

\bal{	% clear
	\label{eq:amplmu}
	\roS[\mu]{[\ta_1,\ta_2]}
	(\psD[\et]{\ta_1} \otimes \coco{\psD[\ki]{\ta_2}}) 
	= \int\!\!\xD \vph_1 \!\! \int\!\!\xD \vph_2\; 
		\psD[\et]{\ta_1}(\vph_1)\,
		\coco{\psD[\ki]{\ta_2}(\vph_2)} \, 
		Z^\mu_{[\ta_1,\ta_2]}(\vph_1,\vph_2).
}
Using \eqref{eq:propmu} and introducing the complex functions $\et^\mu$ and $\ki^\mu$ defined as
\bal{		% revised 2016 APR 07
	\et^\mu(\vc x ) 
	& \defeq \et(\vc x ) + \tfrac{\im}{w(\vc x)}
		\left(\coco{\Up(\ta_1)}\, \mu_1 \right)(\vc x ), 
		&
	% revised 2016 APR 07
	\ki^\mu(\vc x ) 
	& \defeq \ki(\vc x )  - \tfrac{\im}{w(\vc x)}
	\left(\coco{\Up(\ta_2)}\, \mu_2 \right)(\vc x ),
	}
the amplitude \eqref{eq:amplmu} can be expressed in terms of the free amplitude of the coherent states defined by $\et^\mu$ and $\ki^\mu$:
\bal{	% revised 2016 APR 07
	\label{eq:amplmu2}
	\roS[\mu]{[\ta_1,\ta_2]}
	\biglrr{\psD[\et]{\ta_1} \otimes \coco{\psD[\ki]{\ta_2}}}
	= 
	\roS[0]{[\ta_1,\ta_2]}
	\biglrr{\psD[\et^\mu]{\ta_1} \otimes \coco{\psD[\ki^\mu]{\ta_2}}}
	\fracw{\Norm[D,\et]{\ta_1}\, \coco{\Norm[D,\ki]{\ta_2}}}
			{\Norm[D,\et^\mu]{\ta_1}\, \coco{\Norm[D,\ki^\mu]{\ta_2}}}
	\frac{\Norm[Z,\mu]{[\ta_1,\ta_2]}}{\Norm[Z,0]{[\ta_1,\ta_2]}}.
}
Substituting the free amplitude \eqref{zzz_SKG_SF_free_amplitude_11} and the coherent states' normalization \eqref{zzz_SKG_SF_coherent_310}, after a lengthy calculation we arrive at 

\bal{		% revised 2016 APR 13
	\roS[\mu]{[\ta_1,\ta_2]}
	\biglrr{\psD[\et]{\ta_1} \otimes \coco{\psD[\ki]{\ta_2}}}
	&
	 = \roS[0]{[\ta_1,\ta_2]}
	\biglrr{\psD[\et]{\ta_1} \otimes \coco{\psD[\ki]{\ta_2}}}
	\fracw{\Norm[Z,\mu]{[\ta_1,\ta_2]}}{\Norm[Z,0]{[\ta_1,\ta_2]}}
	\exp \biiiglrr{\im \int_{M\lintval\ta12}\!\!\!\!\!\! \xd^4 x
					\sqrt{|g(x)|}\;	\la\htx D_{12}(x)\, \mu(x)\!
					}\quad
			\notag
			\\
	\label{eq:amplmuinterval}
	& \hspace{42mm}
	\times \exp \biiiglrr{\frac{\im}{2} \int_{M\lintval\ta12}\!\!\!\!\!\!
							\xd^4 x \sqrt{|g(x)|}\; \be(x)\, \mu(x)\!
							},
	}
wherein $\be$ results to be
\bal{		% revised 2016 APR 13
	\be(\ta, \vc x ) 
	= \si\int_{\ta_1}^{\ta_2} \xd \ta' 
	\fracw{\sqrt{|g(\ta',\vc x )|}}{w(\vc x) \ti w(\ta) \mc W(\ta)}
	\biiiglrr{\frac{\coco{\Up(\ta)}\,\Up(\ta')}
						{2\im\, \Impart(\coco{c^a}c^b)}
			+ \frac{\mc W^{00}(\ta_1,\ta)\,\mc W^{00}(\ta',\ta_2)}
						{\mc W^{00}(\ta_1,\ta_2)}
			}
	\mu(\ta',\vc x ),
	}
and the complexified classical solution $\la\htx D_{12}$ is
\bal{		% revised 2016 APR 13
	\label{eq:xiinterval}
	\la\htx D_{12}(\ta,\vc x) 
	& = \KD \hat\la_{12}(\ta,\vc x),
		\\
	\hat\la_{12}(\ta,\vc x)
	& = \biglrr{\Up(\ta)\,\et(\vc x)\!
					+\!\coco{\Up(\ta)}\:\coco{\ki(\vc x)}
					}\,
	= \int\!\! \xd^3 k\, \biiglrr{
					\et_{\vc k} \Up_{\!\vc k} (\ta)\, U_{\vc k}(\vc x)
					+\coco{\ki_{\vc k}}\: \coco{\Up_{\!\vc k} (\ta)}\:
						\coco{U_{\vc k}(\vc x)}
					}.	
	}
Our $\hat\la_{12}$ is the $\hat\xi$ defined e.g.~in (81) of \cite{Oe:fermi},
whereas the $\hat\et$ in (39) of \cite{CoOe:smatrix} is our $\la\htx D_{12}$.
This difference is merely a rescaling, which could be removed by redefining the coherent state \eqref{zzz_SKG_SF_coherent_100} in a suitable way.
Substituting expression \eqref{eq:normmu} into\eqref{eq:amplmuinterval}, we obtain the amplitude
\bal{		% revised 2016 APR 13
	\roS[\mu]{[\ta_1,\ta_2]}
	\biglrr{\psD[\et]{\ta_1} \otimes \coco{\psD[\ki]{\ta_2}}}
	= &\; \roS[0]{[\ta_1,\ta_2]}
	\biglrr{\psD[\et]{\ta_1} \otimes \coco{\psD[\ki]{\ta_2}}}
	\exp \biiiglrr{ \int_{M\lintval\ta12}\!\!\!\!\!\! \xd^4 x 
					\sqrt{|g(x)|}\; \la\htx D_{12}(x) \, \mu(x)
					}
		\notag
		\\
	\label{eq:amplmuinterval2}
	& 	\;\times \exp \biiiglrr{ \frac{\im}{2} 
		\int_{M\lintval\ta12}\!\!\!\!\!\! \xd^4 x 
		\int_{M\lintval\ta12}\!\!\!\!\!\! \xd^4 x' 
		\sqrt{|g(x)g(x')|}\; \mu(x) \, G\ltx F(x,x') \, \mu(x') 
		},
	}
in whose last factor $G\ltx F$ arises from $\al+\be$ and writes as
\bal{		% revised 2016 APR 13
	\label{eq:GF}
	G\ltx F(x,x')
	=\KD
		\biiglrr{\te(\ta-\ta')\,\Up(\ta)\, \coco{\Up(\ta')}
				+ \te(\ta'-\ta)\,\Up(\ta')\, \coco{\Up(\ta)}
				}  \tfracw{\im}{w(\vc x)} \de^{(3)}(\vc x\!-\!\vc x').
	}
%
%In bra-ket notation we have again $\roS[\mu]{\intval\ta12} 
%\biglrr{\psD[\et]{\Si\ta_1}\!\otimes\! \coco{\psD[\ze]{\Si\ta_2}}\,} =
%\braopkett{\psD[\ze]{\Si\ta_1}}{\mc S^\mu}{\psD[\et]{\Si\ta_2}}$.
$G\ltx F$ satisfies the inhomogeneous Klein-Gordon equation in both variables $x$ and $x'$, that is $(\si_{00}\Box_x+m^2)\, G\ltx F(x,x')=\de^{(4)}(x-x')/\sqrt{|g(x)|}$.
This can be checked directly 
%by viewing $\im \de^{(3)}(\vc x\!-\!\vc x')/w(\vc x)$ as reduced initial data and then $\Up(\ta)\im \de^{(3)}(\vc x\!-\!\vc x')/w(\vc x)$ being a homogeneous solution, or 
by expanding the Dirac delta using \eqref{eq:modenormbasis}. 
For an interval region in Minkowski spacetime with $\ta=t$ the Minkowski time, $G\ltx F$ coincides with the standard Feynman propagator \cite{CoOe:letter,CoOe:smatrix}, whose familiar form arises from expanding the Dirac delta with \eqref{eq:modenormbasis}. 
The result of this can be found e.g.~in Eq.~(13-30) of Hatfield's QFT book \cite{Hat}. 
The same happens in de Sitter space with $\ta$ equal to the de Sitter conformal time \cite{Co:letter,Co:dS}. 
This justifies our suggestive notation $G\ltx F$: for time-interval regions, \eqref{eq:GF} is the Feynman propagator.

On the right hand side of \eqref{eq:GF} we can see that $\Up$ always appears with the larger foliation parameter as its argument, whereas $\coco\Up$ appears with the lower one.
Hence for time-interval regions we interpret the modes $\Up\lvc k(\ta)\, U\lvc k(\vc x)$ as positive frequency modes, and $\coco{\Up\lvc k(\ta)}\, U\lvc k(\vc x)$ as negative frequency.
For regions where $\ta$ is a spatial coordinate, the modes $\Up\lvc k(\ta)\, U\lvc k(\vc x)$ move in positive $\ta$-direction, whereas $\coco{\Up\lvc k(\ta)}\, U\lvc k(\vc x)$ move in negative $\ta$-direction.
Despite not having specified properties of $X\htx{a}\lvc k(\ta)$, $X\htx{b}\lvc k(\ta)$ and $c\htx{a}\lvc k$, $c\htx{b}\lvc k$ in \eqref{eq:upsilon_k}, this distinction between $\Up$ and $\coco\Up$ is rooted in the positivity condition \eqref{zzz_SKG_SF_vacuum_102}.

We note that all quantities in the interval amplitude \eqref{eq:amplmuinterval2} are well defined, and that it is independent of $\ta_1$ and $\ta_2$, making its limit for asymptotic values of $\ta_1$ and $\ta_2$ trivial.
This justifies its interpretation as the S-matrix for the scalar theory in the presence of a source field.
%
% ==========================================================
%
\subsection{Source field on rod region}
\label{sec:intamplhyp}
\noindent
We consider now the interacting theory in the hypercylinder region. 
The field propagator takes the form

\bal{
	% revised 2016 APR 05
	\label{eq:propmuhyp}
	Z^\mu_R (\vph) 
	& = \fracw{\Norm[Z,\mu]{R}}{\Norm[Z,0]{R}} Z^0_R (\vph)\, 
		\exp \biiiglrr{ \im \int\!\! \xd t \, \xd \Omega \; 
							\mu_R(t, \Omega) \, \vph_R(t, \Omega) 
							},
		\\
	% revised 2016 APR 05
	\mu_R(t, \Omega)
	& = \int_0^R \!\!\xd r \, \sqrt{|g(t, r, \Omega)|} \; 
		\tfracw{X\htx a(r)}{X\htx a(R)} \mu(t,r,\Om).
	}
For $\Norm[Z,\mu]{R}/\Norm[Z,0]{R}$ we get as above
\bal{
	% revised 2016 APR 05
	\label{eq:normmu_rod}
	\frac{\Norm[Z,\mu]{R}}{\Norm[Z,0]{R}} 
	= \exp \biiglrr{ \frac{\im}{2} \int\! \xd^4 x \, 
						\sqrt{|g(x)|}\, \al_R(x) \, \mu(x) 
						},
}
wherein $(\si_{00}\Box+m^2)\,\al_R(x) = \mu(x)$ while $\al_R(t,R,\Om)=0$, and thus $S^\mu_R(\ph+\al_R) = S^0_R(\ph) + \tfrac12 \int_M \xd^4x\,\sqrt{|g|}\,\al_R\, \mu$.
We choose the following real, inhomogeneous solution:
%(we are free to add any homogeneous Klein-Gordon solution $\ti\al_R$ vanishing at $R$):

%
\bal{		% revised 2016 APR 06
	\label{eq:_alphaR}
	\al_R(t,r,\Om) 
	& = \int_0^R\!\! \xd r' \, \sqrt{|g(t, r', \Om)|}
	\fracw{\ka(r,r')}{w(t,\Om) \ti w(r) \mc W(r)} \, 
	\mu(r',t,\Omega),
		\\
	% revised 2016 APR 06
%	% equivalent symmetric form of $\alpha$, left out for brevity
%	\ka(r,r') 
%	& = -\te(r-r')\, X\htx a(r')\, X\htx b(r)
%		- \te(r'-r)\, X\htx a(r)\, X\htx b(r') 
%		+ X\htx a(r)\,X\htx a(r') \tfracw{X\htx b(R)}{X\htx a(R)}.
%		\notag
%		\\
	% revised 2016 APR 06
	\ka(r,r') 
	& = \te(r-r')\, \mc W^{00}(r,r') - X\htx a(r)\, X\htx b(r') 
		+ X\htx a(r)\,X\htx a(r') \tfracw{X\htx b(R)}{X\htx a(R)}.
		\notag
	}
%
%Checking that $\al_R$ vanishes at $R$ is straightforward.
%For verifying that it is an inhomogeneous solution, note that $\Box$ is sensitive to $r$, not to $r'$.
%Hence we can view $\mu(t,r',\Om)$ as reduced initial data, and thus $X\htx{a,b}(r)\,\mu(t,r',\Om)$ are homogeneous solutions, and so are linear combinations of them.
%Acting with the Klein-Gordon operator on \eqref{eq:_alphaR}, it is the $\te$-term which generates the $\mu$-source, while the other terms are homogeneous (since $\ti w(r)\mc W(r)$ is independent of $r$).
%
We now apply the propagator \eqref{eq:propmuhyp} to calculate the amplitude for the boundary state $\coco{\psD[\ki]{R}}$ in the presence of the source $\mu$:

\bal{	% revised 2016 APR 07
	\label{eq:amplsrchyp}
	\roS[\mu]{R}(\coco{\psD[\ki]{R}}) 
	= \int\!\! \xD \vph_R\; 
		\coco{\psD[\ki]{R}(\vph_R)}\, Z^\mu_R (\vph_R).
}
As for interval regions, we introduce the quantity $\ki^\mu_R$
\bal{
	% revised 2016 APR 07
	\ki^\mu_R(t,\Omega) 
	= \ki(t,\Omega) - \tfrac{\im}{w(t,\Om)} 
	\left( \coco{\Up(R)}\, \mu_R \right) (t, \Omega),
}
and then the amplitude \eqref{eq:amplsrchyp} can be expressed in terms of the free amplitude for the coherent state defined by $\ki^\mu_R$:
\bal{		% revised 2016 APR 07
	\label{eq:amplmuhyp2}
	\roS[\mu]{R}\biglrr{\coco{\psD[\ki]{R}}}
	= \roS[0]{R}\biglrr{\coco{\psD[\ki^\mu_R]{R}}}
		\fracw{\coco{\Norm[D,\ki]{R}}}{\coco{\Norm[D,\ki^\mu_R]{R}}}
	\frac{\Norm[Z,\mu]{R}}{\Norm[Z,0]{R}}.
}
Substituting the free amplitude \eqref{zzz_SKG_free_amplitude_100} and the coherent states' normalization \eqref{zzz_SKG_SF_coherent_310}, we obtain 

\bal{		% revised 2016 APR 13
	\roS[\mu]{R}\biglrr{\coco{\psD[\ki]{R}}}
	& = \roS[0]{R}\biglrr{\coco{\psD[\ki]{R}}} \, 
	\frac{\Norm[Z,\mu]{R}}{\Norm[Z,0]{R}} \, 
	\exp \biiiglrr{\im \int_{M_R}\!\!\!\! \xd^4 x
					\sqrt{|g(x)|}\;	\la\htx D_R(x)\, \mu(x) 
					} 
		\notag
			\\
	\label{eq:amplmuhyp3}
	& \hspace{20mm}
	\times \exp \biiiglrr{\frac{\im}{2} \int_{M_R}\!\!\!\!
							\xd^4 x \sqrt{|g(x)|}\; \be_R(x)\, \mu(x)
							},
	}
wherein $\be_R$ results to be
\bal{		% revised 2016 APR 13
	\be_R(\ta, \vc x ) 
	= -\int_{0}^{R} \xd r' 
	\fracw{\sqrt{|g(t,r',\Om)|}}{w(t,\Om) \ti w(r) \mc W(r)}
	\frac{\Up(R)}{c^b}
	\fracw{X^a(r)\,X^a(r')}{X^a(R)}
	\mu(t,r',\Om ),
	}
and the complexified classical solution $\la\htx D_R$ is

\bal{		% revised 2016 APR 13
	\label{eq:xihyp}
	\la\htx D_R(t,r,\Om) 
	& = \frac{\im}{c^b} \fracw{X^a(r)}{\ti w(R) \mc W(R)}
			\coco{\ki(t,\Om)}\,
	=\, \mc K\htx D\, \hat\la_R(t,r,\Om),
		\\
	\hat\la_R(t,r,\Om)
	& = \tfrac{X\htx a(r)\,2\im\,\Impart(\coco{c\htx a}c\htx b)}
					{c\htx b}\:
		\coco{\ki(t,\Om)}.
	}
Our $\hat\la_R$ is the $\hat\xi$ defined e.g.~in (81) of \cite{Oe:fermi},
whereas the $\hat\xi$ in (124) of \cite{CoOe:smatrix} is our $\la\htx D_R$.
Using expression \eqref{eq:normmu_rod}, we obtain the rod amplitude:

\bal{		% revised 2016 APR 13
	\roS[\mu]{R}\biglrr{\coco{\psD[\ki]{R}}}
	& = \roS[0]{R}\biglrr{\coco{\psD[\ki]{R}}}
	\exp \biiiglrr{ \int_{M_R}\!\!\!\! \xd^4 x 
					\sqrt{|g(x)|}\; \la\htx D_R(x) \, \mu(x)
					}
		\notag
		\\
	\label{eq:amplmuhyp4}
	& \hspace{10mm}
	\times \exp \biiiglrr{ \frac{\im}{2} 
		\int_{M_R}\!\!\!\! \xd^4 x 
		\int_{M_R}\!\!\!\! \xd^4 x' 
		\sqrt{|g(x)g(x')|}\; \mu(x) \, G^R\ltx F(x,x') \, \mu(x') 
		},
	}
in whose last factor $G^R\ltx F$ arises from $\al_R \!+\! \be_R$ and writes as
\bal{		% revised 2016 APR 13
	\label{eq:GFhyp}
	G^R\ltx F(x,x')
	= 	\biiglrr{\te(r-r')\,\Up(r)\,X^a(r')
				+ \te(r'-r)\,\Up(r')\,X^a(r)
				} \fracw{-\de(t-t')\,\de^{(2)}(\Om,\Om')}
				{w(t,\Om)\, \ti w(R) \mc W(R)\, c^b}
	.
	}
The propagator $G^R\ltx F$ satisfies the inhomogeneous Klein-Gordon equation in both variables $x$ and $x'$, that is $(\si_{00}\Box_x+m^2)\, G\ltx F(x,x')=\de^{(4)}(x-x')/\sqrt{|g(x)|}$.
This can be checked in the same way as for the interval region.
On the right hand side of \eqref{eq:GFhyp} we can see that $\Up$ always appears with the larger radius as its argument, whereas $X\htx a$ appears with the lower radius.
Hence we interpret the modes $\Up_{\om l m_l\!}(r)\, U_{\om l m_l\!}(t,\Om)$ as outgoing modes (moving in increasing $r$-direction), and $\coco{\Up_{\om l m_l\!}(r)}\, U_{\om l m_l\!}(t,\Om)$ as incoming (moving in decreasing $r$-direction).
Despite not restricting $\smash{c\htx{a}\lvc k}$ and $\smash{c\htx{a}\lvc k}$ in \eqref{eq:upsilon_k}, this distinction between $\Up$ and $\coco\Up$ is again due to the the positivity condition \eqref{zzz_SKG_SF_vacuum_102}.
Further, we recall that $X\htx a(r)$ represents the modes which are regular for all radii, whereas the $X\htx b (r)$ become singular.
In \eqref{eq:GFhyp} the lower radius always appears in $X\htx a$, so no singularity occurs here, because the $X\htx b$ in $\Up$ has the larger radius as its argument.
Hence all quantitites in the rod amplitude \eqref{eq:amplmuhyp4} are well defined.

The structure of the rod amplitude \eqref{eq:amplmuhyp4} is the same as that of the interval amplitude \eqref{eq:amplmuinterval2}.
The rod amplitude \eqref{eq:amplmuhyp4} is independent of $R$, making the asymptotic limit $R\to\infty$ is trivial again. 
Hence, we can interpret it as the radial S-matrix for the scalar field theory in the presence of a source.
%
% ==============================================================
%
\subsection{General interaction}
\label{sec:genint}
\noindent
The asymptotic amplitude for a general interacting theory (labeled by $V$) can be worked out perturbatively applying functional derivatives.
The action of the scalar field with an arbitrary potential $V$ in a spacetime region $M$ is given by
\bal{		% clear
	S^V_{M}(\phi) 
	= S^0_M(\phi) + \int_M \! \xd^4 x \sqrt{|g(x)|}\; V(x,\phi(x)).
	}
We can write $\exp (\im S^V_{M}(\phi))$ as an infinite series of variational operators acting on the corresponding term in the presence of a source field
\bal{		% clear
\exp ( \im S^V_{M}(\phi)) 
	= \exp \biiglrr{ \im \int_M \!\xd^4 x \sqrt{|g(x)|}\;
									V(x,-\im \tfrac{\delta}{\delta \mu(x)})
									} 
	\exp (\im S^\mu_M(\phi)) \biigr|_{\mu=0},
	}
wherein $S^\mu_M$ is the action \eqref{eq:actionsrc} for a source interaction. 
We assume that the potential $V$ vanishes outside of the region $M$.
Inserting the above expression in the field propagator \eqref{eq:fieldprop} leads to 
\bal{		% clear
	Z^V_M(\vph) 
	= \exp \biiglrr{ \im \int_M\! \xd^4 x \sqrt{|g(x)|}\; 
			V(x, -\im \tfrac{\delta}{\delta\mu(x)})
			}\,
	Z^\mu_M(\vph)\biigr|_{\mu=0},
	}
which lets the amplitude for the general interacting theory become
(for interval and rod regions)
\bal{		% clear
	\roS[V]M(\vph) 
	= \exp \biiglrr{ \im \int_M\! \xd^4 x \sqrt{|g(x)|}\; 
			V(x, -\im \tfrac{\delta}{\delta\mu(x)})
			}\,
	\roS[\mu]M(\vph)\biigr|_{\mu=0}.
	}
%
% =======================================================
%
\section{Examples}
\label{sec:examples}
\noindent
In this section we provide some examples showing the consistency of our general expressions with some results obtained in Minkowski and de Sitter spacetimes. 
In particular, we indicate the main operators involved and also the mode decompositions from which the vacuum state, amplitudes and Feynman propagators can be recovered.
%
%========================================================
%
\subsection{Interval region in Minkowski and de Sitter spacetimes}
\label{sec:intervalMindS}
\noindent
As a first example, we consider the standard time-interval regions in Minkowski spacetime: The foliation parameter $\ta$ coincides with the global time variable $t$, and the three coordinates $\vc x $ are the usual cartesian spatial coordinates. 
This is the standard situation with spacetime foliated by equal-time hyperplanes. 
Plane waves form a useful orthonormal basis to expand the boundary field configurations: $U_{\vc k}(\vc x) =  \eu^{\im \vc k\, \vc x} (2\pi)^{-3}/(2E_{\vc k})$.
We can choose $X\htx a(\ta) = \cos(\omega \ta)$ and $X\htx b(\ta) = \sin (\omega \ta)$, wherein $\omega \defeq \sqrt{-\Delta_{\vc x } + m^2}$ with $\Delta_{\vc x }$ denoting the Laplacian in the coordinates $\vc x$. 
For the interval region $M\lintval t12$, the matrix $W_{[t_1,t_2]}$ becomes
\bals{W_{[t_1,t_2]} 
	= \frac{\omega}{\sin\biglrr{\omega(t_2\!-\!t_1)} }
	\begin{pmatrix}
		\cos\biglrr{\omega(t_2\!-\!t_1)} 
		& -1 
			\\
		-1 
		& \cos\biglrr{\omega(t_2\!-\!t_1)}
	\end{pmatrix}.
	}
The choice $c\htx a = (2\pi)^3\, 2E_{\vc k}$ and $c\htx b = -\im\,(2\pi)^3\, 2E_{\vc k}$ then induces the usual vacuum state with $A_{\Si_t}=\om$,
which coincides with formula (15) of \cite{CoOe:smatrix}.

In the case of a massive scalar field in de Sitter space, the time-interval region considered in \cite{Co:dS} is bounded by two hypersurfaces of constant conformal de Sitter time $t$. 
As in Minkowski spacetime, the modes $U_{\vc k }(\vc x )$ are plane waves, $\eu^{\im \vc k  \, \vc x }$. 
Now, the operators $X\htx{a}$ and $X\htx{b}$ arise from a Bessel equation: $X\htx a(t) = t^{3/2} J_{\nu}(k t)$ and $X\htx b(t) = t^{3/2} Y_{\nu}(k t)$ wherein $ k = |\vc k |$.
$J_{\nu}$ and $Y_{\nu}$ are the Bessel functions of the first and second kind respectively, with index $\nu=\sqrt{9/4 - (mR)^2}$ and $R$ denotes the inverse of the Hubble constant. 
For the time-interval region $[t_1,t_2]$ the elements of the matrix $W_{[t_1,t_2]}$ are
\bal{
	W_{[t_1, t_2]}^{11} 
	&= -\frac{R^2}{t_1^2} 
		\left( \frac{3}{2 t_1} 
			+ k \frac{J_{\nu}'(k t_1)\,Y_{\nu}(k t_2) 
						- Y_{\nu}'(k t_1) \, J_{\nu}(k t_2)}
						{J_{\nu}(k t_1) \, Y_{\nu}(k t_2) 
						- Y_{\nu}(k t_1) \, J_{\nu}(k t_2)}
				\right), 
		\\
	W_{[t_1, t_2]}^{12} 
	&= W_{[t_1, t_2]}^{21} 
	= \frac{-2\, R^2\,(t_1 t_2)^{-3/2}/\pi}
			{J_{\nu}(k t_1) \, Y_{\nu}(k t_2) 
			- Y_{\nu}(k t_1) \, J_{\nu}(k t_2)},
		\\
	W_{[t_1, t_2]}^{22} 
	&= \frac{R^2}{t_2^2} 
		\left( \frac{3}{2 t_2} 
			+ k \frac{J_{\nu}(k t_1) \, Y_{\nu}'(k t_2) 
						- Y_{\nu}(k t_1) \, J_{\nu}'(k t_2)}
						{J_{\nu}(k t_1) \, Y_{\nu}(k t_2) 
						- Y_{\nu}(k t_1) \, J_{\nu}(k t_2)}
			\right),
	}
wherein a prime indicates the derivative with respect to the argument. 
The vacuum state is obtained again by fixing $c\htx a=1$ and $c\htx b=\im$.

Inserting all the above formulas in the free amplitude \eqref{zzz_SKG_SF_free_amplitude_11}, the complex function $\hat\la$ \eqref{eq:xiinterval} and the Feynman propagator \eqref{eq:GF} provides the correct expressions for the corresponding quantities in Minkowski \cite{CoOe:smatrix} and de Sitter spacetimes \cite{Co:dS}.
%
% ==========================================================
%
\subsection{Hypercylinder region in Minkowski and de Sitter spacetimes}
\label{sec:hypMindS}
\noindent
In Minkowski spacetime, the modes $U_{E l m_l}(t,\Om)$ are the product of spherical harmonics $Y_l^{m_l}(\Om)$ and the exponential $(2 \pi)^{-1/2} \eu^{-\im E t}$, with $E \in \reals$. 
We set $X\htx a(r) \!=\! a_l(E,r)$ and $X\htx b(r) \!=\! b_l(E,r)$, wherein
\bal{
	a_l(E, r) 
	& = \begin{cases} 
			j_l(r \sqrt{E^2 - m^2}) & \text{if } E^2>m^2 
				\\ 
			i^+_l(r \sqrt{m^2 -E^2}) & \text{if } E^2<m^2
		\end{cases},
		&
	b_l(E, r) 
	& = \begin{cases} 
			n_l(r \sqrt{E^2 - m^2}) & \text{if } E^2>m^2
				\\ 
			i^-_l(r \sqrt{m^2 -E^2}) & \text{if } E^2<m^2
		\end{cases}.
	}
$j_l$ and $n_l$ are the spherical Bessel functions of the first and second kind, and $i^+_l$ and $i^-_l$ are the modified spherical Bessel functions of the first and second kind. 
In the region between two hypercylinders of radii $R_1$ and $R2$, the matrix $W_{[R_1,R_2]}$ has the form
\bal{W_{[R_1,R_2]} 
	= \frac{1}{ \mc W^{00}(R_1,R_2)}
		\begin{pmatrix}   
		R_1^2\, \mc W^{10}(R_1,R_2) & 1/p 
			\\ 
		1/p & - R_2^2\, \mc W^{01}(R_1,R_2) 
		\end{pmatrix},
	}
wherein $\mc W^{jk}$ are defined as in \eqref{eq:wron-components}. 
The vacuum state is fixed by the choice $c\htx a=1$ and $c\htx b=\im$.

In de Sitter space, the modes are $U_{k l m_l}(t, \Om) = t^{3/2} H_{\nu}(k t) Y_l^{m_l}(\Om)$, wherein $H_{\nu}$ is the MacDonald function of index $\nu$.
The operators $X\htx{a}$ and $X\htx{b}$ are given by the spherical Bessel functions of the first and second kind only: $X\htx a(r) = j_l(k r)$ and $X\htx b(r) = n_l(k r)$.
The matrix $W_{[R_1,R_2]}$ reads
\bal{W_{[R_1,R_2]} 
	= \frac{R^2}{t^2} \frac{1}{\mc W^{00}(R_1,R_2)}
		\begin{pmatrix}  
		- R_1^2\, \mc W^{10}(R_1,R_2) & 1/k 
			\\ 
		1/k & R_2^2\, \mc W^{01}(R_1,R_2) 
		\end{pmatrix}.
	}
Our choice for the vacuum state corresponds to $c\htx a=1$ and $c\htx b=\im$.

With all these expressions at our disposal we can immediately obtain the free amplitude \eqref{zzz_SKG_SF_free_amplitude_11_tube}, the complex function $\hat\la$ \eqref{eq:xiinterval} and the Feynman propagator \eqref{eq:GF}, that coincide with those evaluated in \cite{CoOe:smatrix} and \cite{Co:dS}.
%
% =================================================================
%
\section{Summary and outlook}
\label{sec:discussion}
\noindent
In this article we have implemented the quantization of a real, massive, 
scalar field in a 4-dimensional curved spacetime according 
to the prescriptions of the General Boundary Formulation (GBF) 
of Quantum Field Theory (QFT). 
We consider spacetimes that admit at least one global time coordinate,
and also have foliations for which the metric becomes block-diagonal
with respect to the foliation parameter,
see Sections~\ref{sec:interval} and \ref{sec:hypercylinder}.
Although this assumption seems very restrictive, it is nevertheless satisfied 
for many spacetimes in which QFTs have been studied so far,
in particular for all globally hyperbolic spacetimes
\cite{BernalSanchez:_MetricSplitting}, 
for Anti de Sitter, and for black hole spacetimes (Kerr-Newmann).
We emphasize that these foliations may refer to the same spacetime, and are useful
to define two types of spacetime regions characterized by different boundaries.
We also require that using the chosen foliation we can apply separation of
variables to solve the Klein-Gordon equation. 

The first type of regions, called interval regions, is bounded by two hypersurfaces (not required to be Cauchy surfaces or spacelike hypersurfaces).
By contrast, the boundary of the second kind, called rod region, consists of only one connected hypersurface, which is called a hypercylinder. 
Moreover, we assume that the solutions of the Klein-Gordon equation in the different regions can be written in a special form, i.e. in terms of operators satisfying certain conditions specified in Section \ref{sec:interval} and Section \ref{sec:hypercylinder}.

Our main objective is the derivation of an explicit expression for the S-matrix in the case of a general interacting theory. 
This has been achieved by constructing the relevant quantities: The state spaces associated to the boundaries of the interval and hypercylinder regions, the field propagators encoding the dynamics of the field in these regions, the vacuum state and the coherent states. 
Applying then the procedure of \cite{CoOe:letter,CoOe:smatrix}, we compute the amplitude for boundary coherent states in the regions considered for three cases: first for the free theory in the interaction picture, second for an interaction with a source field, and third the general interacting theory (using functional derivative techniques). 
The asymptotic limit of these amplitudes can be interpreted as the S-matrices for the scalar field defined in the two types of regions. 

The structures of these S-matrices for interval and rod regions are similar. 
In the case of the source interaction, the asymptotic amplitude for coherent states factors into three terms: 
First, the amplitude of the free theory, second, an exponential coupling the source to a function which establishes a one-to-one correspondence between complex solutions of the equation of motion and coherent states, and third, a term bilinear in the source in which the Feynman propagator appears. 
In the case of an interacting scalar field in Minkowski spacetime studied in \cite{CoOe:letter,CoOe:smatrix}, the Feynman propagator obtained in the interval region and the one in the rod region were shown to be equivalent. 
Equating the complex function $\hat\xi$ appearing in (the corresponding formulas of ours) \eqref{eq:amplmuinterval2} and \eqref{eq:amplmuhyp2}, an isomorphism was constructed between the state spaces associated to the boundaries of the two regions.
Then the equivalence of the free amplitudes under the action of this isomorphism  was shown. 
In this way the interacting theory in the rod region turned out to provide the same asymptotic amplitudes as the standard treatment based on the time interval region. 
An analogous result was recovered in de Sitter space \cite{Co:letter,Co:dS} and Rindler space \cite{CoRa:rindler}. 
A natural question would then be if a similar situation is to be expected here. 
To address this question, that we expect should be answered in the positive at least for field theories defined on spacetimes conformal to Minkowski spacetime, many strategies (not all independent) can be envisaged: defining a map from the space of classical solutions of the equation of motion in the interval region to the space of solutions in the whole spacetime, and a second map from this space to the space of solutions in the rod region and then composing these two maps to relate solutions in the two regions of interest.
This implements a Bogolubov transformation between the two sets of modes in which the field $\phi$ has been expanded.
Related strategies are finding a relation between the vacuum states defined on the boundary of the two regions or showing the equivalence of the different boundary conditions satisfied by the Feynman propagators of interval and rod regions. 
However, so far no general result has been obtained along these lines.

Other important aspects that deserve to be investigated concern the analytic properties and the unitarity of the S-matrix. 
A first step in the study of unitary of quantum dynamics of a scalar field in curved spacetime within the GBF was taken in \cite{CoOe:unit}, where results were obtained for the free theory as well as in the presence of a source interaction. 
The technique used in that paper is based on a the composition property of the field propagator. 
With the general structure of the S-matrix at our disposal, this question can now be handled from a new perspective.

There are other directions of generalizing the work presented here. 
The first natural extension would be to compute the S-matrix for fields defined in more general regions. 
In particular, compact spacetime regions will play a major role, e.g., causal diamonds and 4-balls. 
Second, more interesting field theories need to be considered: Fields of higher spin have to be taken into account in order to investigate QED or Yang-Mills Theory from a GBF perspective
(a first step in this direction was taken in \cite{Oe:2dqym}).

Besides its contribution to the development of the GBF, our result should also be useful in the study of QFTs in spacetimes with boundaries, e.g., mirrors or horizons, that appear in many contexts \cite{BiDa:qftcurved}, like the Casimir effect \cite{Cas,Bor,Mil,Dal}. 
Moreover, considering the hypercylinder regions enables us to define S-matrices for spacetimes where this has not been possible in the standard formalism due to the lack of free temporal asymptotic states, such as Anti-de Sitter (AdS) spacetime (rod region) or in the case of a field in the presence of an eternal black hole (tube region). 
A boundary S-matrix in AdS for states defined at timelike infinity (which plays an important role in the conjectured AdS/CFT correspondence \cite{Ma:AdS_CFT}) was proposed in \cite{Balasubramanian:1999ri,Giddings:1999qu}. 
The GBF is likely to provide the appropriate tool for a rigorous derivation of this type of S-matrix. 
In the same spirit the hypercylinder geometry appears to be the necessary ingredient in 't Hooft's proposal \cite{tHooft:1996tq} to describe the scattering of particles against a black hole. 
Therefore a next step will be the application of our work to these systems.

The expressions found here for the Feynman propagator in certain curved spacetimes enable us to also write down the other members of the Green functions family, in particular the Wightman, Schwinger and Hadamard functions.
Thus we are in a position where we can (at least formally) define canonical commutation relations (CCR) for field operators on general hypersurfaces. 
If and which physical information is contained in these CCR has to be explored.
Based on the CCR, the development of a formulation of canonical quantization within the GBF seems to be possible. 
It will then be interesting to investigate the relations between the GBF and Algebraic QFT.

So far, two methods of quantization have been studied within the GBF framework: The Schr\"odinger-Feynman quantization used e.g.~in this paper and the Holomorphic Quantization developed in \cite{Oe:hol}.
While Schr\"odinger-Feynman quantization is a rather heuristic method,
it can be apllied to interacting theories.
On the other hand, Holomorphic Quantization is mathematically rigorous,
but at present applies to linear/affine field theories only.
A natural extension of the result presented here will be to recover it by applying the 
%In a follow-up article to the present work
%we confirm the results found here: We first apply 
Holomorphic Quantization, and to relate it
%then relate its results
to the Schr\"odinger-Feynman one using the general correspondence
between the two representations found in \cite{Oe:Sch-hol}.
We hope that in future work these results can shed some light on 
how to include interactions into the holomorphic quantization scheme.
%
% ==============================================================
%
\begin{acknowledgments}
\noindent
The authors are especially grateful to Robert Oeckl (CCM-UNAM Morelia) for 
many helpful discussions and comments on a draft of this article.
This work was supported in part by
UNAM-DGAPA-PAPIIT project grant IA105416 (DC), 
and CONACyT scholarship 213531, 
UNAM-DGAPA-PAPIIT project grant IN100212 
and PRODEP scholarship DSA/103.5/16/4816 of the SEP of Mexico (MD).
\end{acknowledgments}
%\bibliographystyle{unsrt} % order as cited in text
\bibliographystyle{plain} % order authors alphabetically 
\bibliography{references_S_curved}
%
% ================================================================
%
\end{document}